\documentclass[submission, Phys]{SciPost}
\usepackage{amsmath,amsfonts,amssymb}
\usepackage{bbold}

\begin{document}

\begin{center}{\Large \textbf{
Interedge backscattering in time-reversal symmetric quantum spin Hall Josephson junctions
}}\end{center}

\begin{center}
C. Heinz\textsuperscript{1},
P. Recher\textsuperscript{1,2},
F. Dominguez\textsuperscript{3}
\end{center}

\begin{center}
{\bf 1} Institut f\"ur Mathematische Physik, Technische Universit\"at Braunschweig, D-38106 Braunschweig, Germany
\\
{\bf 2} Laboratory for Emerging Nanometrology Braunschweig, D-38106 Braunschweig, Germany
\\
{\bf 2} Institute for Theoretical Physics and Astrophysics, and W\"urzburg-Dresden Cluster of Excellence on Complexity, Topology and Dynamics in Quantum Matter ctd.qmat, Julius-Maximilians-Universit\"at W\"urzburg, Am Hubland, D-97074 W\"urzburg, Germany
\\

* f.dominguez@uni-wuerzburg.de
\end{center}

\begin{center}
\today
\end{center}

\section*{Abstract}
{\bf
% TODO: write your abstract here.
Using standard tight-binding methods, we investigate a novel backscattering mechanism taking place on quantum spin Hall N'SNSN' Josephson junctions in the presence of time-reversal symmetry. This extended geometry allows for the interplay between two types of Andreev bound states (ABS): the usual phase-dependent ABS localized at the edges of the central SNS junction \emph{and} phase-independent ABS localized at the edges of the N'S regions. Crucially, the latter arise at discrete energies $E_n$ and mediate a backscattering process between opposite edges on the SNS junction, yielding gap openings when both types of ABS are coherently coupled.
In this scenario, a 4$\pi$-periodic ABS decouples from the rest of the 2$\pi$-periodic spectrum, yielding several observable consequences: 
Firstly, we show that the $4\pi$-periodic spectrum can be probed by means of the Shapiro experiment even in the presence of dynamical transitions between the ABS and the quasicontinuum. Secondly, the presence of this backscattering mechanism distorts the superconducting quantum interference (SQI) pattern within the length scale, determined by the ratio between $4\pi$- and $2\pi$-periodic supercurrent contributions. Finally, we propose to use a magnetic flux to tune $E_n$ to zero, resulting in the selective lifting of the fractional Josephson effect.
}

\vspace{10pt}
\noindent\rule{\textwidth}{1pt}
\tableofcontents\thispagestyle{fancy}
\noindent\rule{\textwidth}{1pt}
\vspace{10pt}

\section{Introduction}
\label{sec:intro}
% TODO: write your article here.
Josephson junctions (JJs) based on a single proximitized quantum spin-Hall (QSH) edge~\cite{Fu2009a,Beenakker2013a} reveal their topological character by the presence of zero energy Majorana bound states for the superconducting phase difference $\phi=\pi$. Under parity conservation, the Josephson frequency is halved with respect to the conventional one, $f=2eV/h \rightarrow eV/h$. The so-called \emph{fractional Josephson effect} can, thus, be detected by probing observables linked to the periodicity of the supercurrent, e.g.~the Shapiro experiment, developing constant voltage steps cleaved at $V_n=n \hbar \omega/2e$ ($V_n=n \hbar \omega/e$) for conventional (topological) JJs, with $\omega$ an external frequency, or measuring the Josephson radiation, with frequency $f$ ($f/2$) in conventional (topological) JJs. 

There is however a practical problem that one needs to circumvent to measure the fractional Josephson effect: 
in the presence of time-reversal symmetry (TRS), single-edge ABS are protected against backscattering, yielding a spectrum with no gap openings. Hence, a driven scenario leads irrevocably to an exchange of particles with the quasicontinuum, with the consequent change of parity and the destruction of the 4$\pi$-periodicity. 

Opening sizeable gaps by breaking TRS, i.e.~adding an external magnetic field or magnetic add atoms, is technically difficult and can lead to unwanted phenomena, such as screening currents on the superconductor. 
Indeed, experiments performed so far in quantum spin-Hall Josephson junctions preserve TRS~\cite{Hart2014, Pribiag2015a, Bocquillon2016a, Deacon2017a, Bendias2018, Randle2023}. Unexpectedly, two of these experiments have shown signatures compatible with a topological ground state, with the absence of odd Shapiro steps~\cite{Bocquillon2016a} and the measurement of the fractional Josephson frequency $f/2$ in the Josephson radiation~\cite{Deacon2017a}. Previous theoretical works have made these experimental findings compatible with the absence of an explicit TRS breaking mechanism, by either considering a two-particle backscattering mechanism with large dissipation~\cite{Sticlet2018a} or a retardation effect present both in trivial and in topological superconductors~\cite{Lahiri2023a}. Alternatively, one could attribute the 4$\pi$-periodicity to a trivial scenario with
non-adiabatic transitions between Andreev bound states~\cite{Yeyati2003a,San-Jose2012a,Dominguez2012a,Pikulin2012a, Houzet2013a, Virtanen2013a, Matthews2014a, Sau2017a} or by the presence of an environmental parasitic impedance~\cite{liu2024}.

\begin{figure}[t!]
  \centering
  \includegraphics[width=1\linewidth]{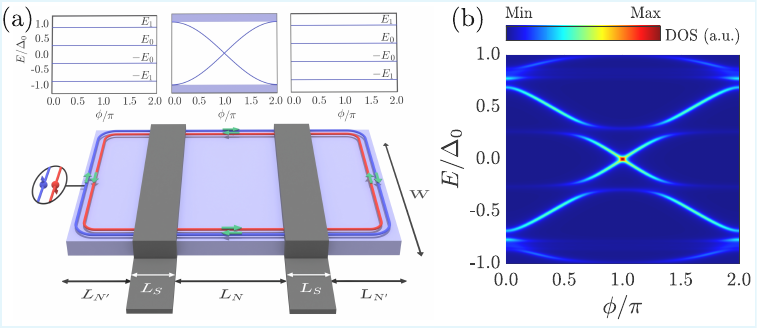}
  \caption{Panel (a): Sketch of the extended quantum spin Hall Josephson junction. Here, the partial cover of the QSH bar (light blue) by superconducting leads (black) defines the extended N'SNSN' Josephson junction. Helical edge states are represented by blue and red curves. 
  Inset: Andreev spectra as a function of the phase difference $\phi$ of the disconnected parts N'S (left), SNS (center) and SN' (right) regions. Panel (b): DOS for a N'SNSN' Josephson junction with $L_s=0.38\,\mu$m and $\xi_s=0.19\,\mu$m, $L_{N'}=0.5\,\mu$m. Avoided crossings coincide with the position of the energy levels at $E\approx 0.25\Delta_0$ and $0.65\Delta_0$.}
  \label{fig:setup}
\end{figure}

In this work, we investigate a way to isolate energetically the topological ABS from the quasicontinuum without breaking time-reversal symmetry.  To this aim, we engineer a backscattering process between opposite edges, mediated by an additional ABS present in the extended N'SNSN' junction, see Fig.~\ref{fig:setup}(a)~\footnote{Alternatively, one can design single-edge backscattering by adding an extra normal part to a JJ, i.e. $\text{S}_1\text{N}\text{S}_2\text{N}^{'} \text{S}_2\text{N}\text{S}_1$, with two superconductors $\text{S}_{1,2}$, differing by a phase difference.}. In this geometry, the central SNS junction is embedded in two additional normal N' parts resulting, for example, from the partial covering of the QSH bar with superconducting fingers, see Fig.~\ref{fig:setup}(a)~\footnote{Note that similar results can be obtained if only one external N' part is present in the setup, see App.~\ref{App: N'SNS -junction}. However, we consider two external parts to remain closer to previous experimental realizations, see~\cite{Bocquillon2016a,Deacon2017a}.}. Here, two types of ABS arise: the usual phase-dependent ABS~\cite{Fu2009a, Beenakker2013a} localized at the edges of the SNS region, and a phase-independent ABS localized at the N'S regions. The latter exhibits a discrete energy spectrum determined by the perimeter of the N' regions, namely
\begin{align}
  E_n=\frac{\hbar v_F }{p_{N'}} \pi (n+1/2),  
\end{align}
with $n\in \mathbb{Z}$ and $p_{N'}=2L_{N'}+\text{W}$, see Fig.~\ref{fig:setup}(a). The coupling between both types of ABS becomes effective when they are in resonance and when the width of the superconducting fingers is smaller or of the order of the superconducting coherence length ($L_s\lesssim \xi_s$). In this scenario, avoided level crossings develop, yielding a 4$\pi$-periodic ABS decoupled from the rest of the spectrum.

In the rest of this contribution, we analyze measurable consequences of the presence of these avoided level crossings. Firstly, we study the Shapiro steps resulting from the modified ratio of the 4$\pi$- and 2$\pi$-periodic critical currents,~$I_{\text{c},4\pi}/I_{\text{c},2\pi}$, allowing for dynamical transitions between them. Here, we observe a suppression of the odd voltage steps, the hallmark of the fractional Josephson effect, under realistic conditions. Secondly, we investigate the distortion of the SQI pattern due to the presence of the additional length scale $L_{N'}$. 

\section{Tight-binding model}
\label{Tight-binding model}
We model the Josephson junction depicted in Fig.~\ref{fig:setup}(a) by means of the proximitized BHZ Hamiltonian~\cite{Bernevig2006a} in the absence of Rashba or Dresselhaus spin-orbit coupling. Even though spin-orbit couplings are naturally present in QSH insulators, such as HgTe \cite{Liu2008a,Murani2017a}, here, we assume they will not modify fundamentally the backscattering mechanism between ABS. This approximation simplifies our numerical calculations since we can express the resulting Bogoliubov de Gennes (BdG) Hamiltonian in a reduced basis, that is, $H=(1/2)\int \text{d}r^2 \Psi^{\dagger}({\bf r})\mathcal{H}\Psi({\bf r})$, with
\begin{align}
\label{eq:BdGBHZ}
    \mathcal{H}=\begin{pmatrix}
        \mathcal{H}_{e} & \Delta \mathbb{1}_{2\times 2}\\
        \Delta^{*}\mathbb{1}_{2\times 2} & -\mathcal{H}_{e}^*
    \end{pmatrix},
\end{align}
and the field operator 
\begin{align}
\label{Electron basis}
    \Psi({\bf r})=\left[c_{E\uparrow}({\bf r}),c_{H\uparrow}({\bf r}),c_{E\downarrow}^{\dagger}({\bf r}),c_{H\downarrow}^{\dagger}({\bf r})\right]^{T},
\end{align}
written in the reduced subspace for BdG electrons (holes) with spin $\uparrow(\downarrow)$. Here, $c_{a\sigma}^{(\dagger)}({\bf r})$, destroys (creates) an electron with orbital $a=E,H$, spin $\sigma=\uparrow,\downarrow$ at position ${\bf r}$. 

The electronic part of the BdG Hamiltonian is given by the BHZ model $\mathcal{H}_{e}=\varepsilon(\hat{k})+M(\hat{k})\sigma_{z}+A\left(\hat{k}_{x}\sigma_{x}-\hat{k}_{y}\sigma_{y}\right)$, with $\varepsilon(\hat{k})=C-D\hat{\textbf{k}}^2$, $M(\hat{k})=M-B\hat{\textbf{k}}^2$ and $\hat{\textbf{k}}=-i\hbar \nabla_{\textbf{r}}$, with the Pauli matrices $\sigma$ operating on the orbital degree of freedom ($E$, $H$) and parameters detailed in the footnote~\footnote{We use $A=373\,$meV\,nm, $B=-857\,$meV, $D=-682\,$meV\,nm$^2$, $M=-10\,$meV and $\Delta_0=0.6\,$meV. We set the chemical potential by means of $C$. On the normal part $C_n=-8\,$meV and the superconducting part $C_s=-20\,$meV.}. In the BHZ model, $M$ opens a gap in the semiconductor spectrum, yielding a trivial insulating gap for $M>0$ and a non-trivial one for $M<0$, with the emergence of helical edge states. 
Moreover, we model the superconducting leads by introducing an energy-independent superconducting pairing $\Delta({\bf r})$, see App.~\ref{App: Energy-independent superconducting pairing}. We take $\Delta({\bf r})$ to be constant in $y$-direction and step-like in the $x$-direction, with $\Delta({\bf r})=\Delta_0 \exp(i s \phi/2)$ within the ranges $L_N/2+L_s\geq |x| \geq L_N/2$ and zero otherwise, with $s=\text{sign}(x)$.

Using standard finite difference methods, we discretize the Hamiltonian~\eqref{eq:BdGBHZ} replacing the coordinate ${\bf r}\rightarrow ( i, j )a$, with $i,j\in \mathbb{Z}$ and the lattice constant $a=5\,$nm. 
Then, we model the dimensions and length scales of the Josephson junction guided by the physical phenomena taking place in typical experimental setups. Namely, a width W so large such that the overlap between opposite edges~\cite{Zhou2008a} and cross-Andreev~\cite{Reinthaler2013a} processes are negligible. To this aim, we use $\text{W}=1\,\mu$m, $M=-10\,$meV and $\Delta_0=0.6\,$meV, yielding a coherence length $\xi_s\approx 190\,$nm and a QSH edge localization length $l_\text{loc}\approx 23\,$nm \cite{Zhou2008a}. Accordingly, we use the superconducting leads thickness of $L_s=380\,$nm, such that the QSH edges placed on N and N' are coupled, see Fig.~\ref{fig:setup}(a). 

\section{Coupled Andreev bound states }
Once we have set the length scales of the system, we move on and study the Andreev bound state spectrum by computing the density of states (DOS) along a section on the normal central part of the Josephson junction as a function of $\phi$. Due to the large size of the discrete system [$\text{dim}(\mathcal{H})\sim 10^5$], we make use of recursive Green's functions (GFs) methods~\cite{MacKinnon1985} and express the DOS in terms of the imaginary part of the advanced GF along the normal part N, namely
\begin{align}
\label{eq:DOS}
    \text{DOS}(E)=\frac{1}{\pi} \text{Im}\sum_{x=1}^{n}\text{Tr}_\text{W}\{G^{a}(x,x,E)\},
\end{align}

with the GFs evaluated at position $x$ and energy $E$, along a segment of the normal part N of length $n a$, where $n=20$~\footnote{The fluctuations in the density of states between different stripes are minimal. Therefore, it is sufficient to restrict the calculation to a small portion of the normal region.}. This GF is represented by a $4 \bar{\text{W}} \times 4 \bar{\text{W}}$ matrix, with $\bar{\text{W}}\equiv \text{W}/a=200$, see further details in App.~\ref{App.transport}.

In Fig.~\ref{fig:setup}(b), we show a representative example of the energy-phase relation for the extended Josephson junction N'SNSN' with $L_s=0.38\,\mu$m. Here, we can observe the emergence of avoided level crossings around the phase-independent ABS positions, i.e.~$E_0\approx 0.25\Delta_0$ and $E_1\approx 0.65\Delta_0$. The size of these gaps scales with the geometric factor $\sim \Delta_0 \exp(- L_s/\xi_s)\approx 0.1 \Delta_0$, which determines the transparency of the superconducting lead~\cite{Finocchiaro2018a}\footnote{Note that disorder in the superconducting regions can affect the transparency. For this setup, however, we have checked onsite disorder strengths up to $\lambda=5\,\mathrm{meV}$ and found no qualitative change in the ABS spectrum, as discussed in detail in App.~\ref{App.disorder}.}. Naturally, in the limit of $L_s\gg \xi_s$, we recover the ABS spectrum touching the quasicontinuum. 

Before analyzing the consequences of the presence of anticrossings, it is worth stressing the differences between the backscattering process studied here and those resulting from either the direct overlap between the QSH wavefunctions along the normal part~\cite{Knapp2020a}, or just at the NS interface~\cite{Lee2014a, Dominguez2024a,Recher2013}. In those cases, the ABS spectrum develops gap openings at every time-reversal symmetric point $\phi=0$ and $\phi=\pi$, see further aspects of this backscattering process in App.~\ref{App:Direct edge coupling}. Only for some specific ``sweet spots" in the parameter regime, one can find crossings at $\phi=\pi$ and avoided level crossings at $\phi=2n\pi$, with $n\in \mathbb{Z}$. However, these ``sweet spots" turn into anticrossings immediately, by changing any parameter of the system like the gate voltage, showing a lack of robustness for any practical purposes. In contrast, in the extended N'SNSN' junction, the gap openings are set by the geometry of the junction, yielding an unperturbed zero energy crossing as long as the discrete level $E_0$ is away from zero energy, i.e.~$E_0\gg \Delta_0 \exp(- L_s/\xi_s)$. 

\begin{figure*}[tb]
  \centering
  \includegraphics[width=1\linewidth]{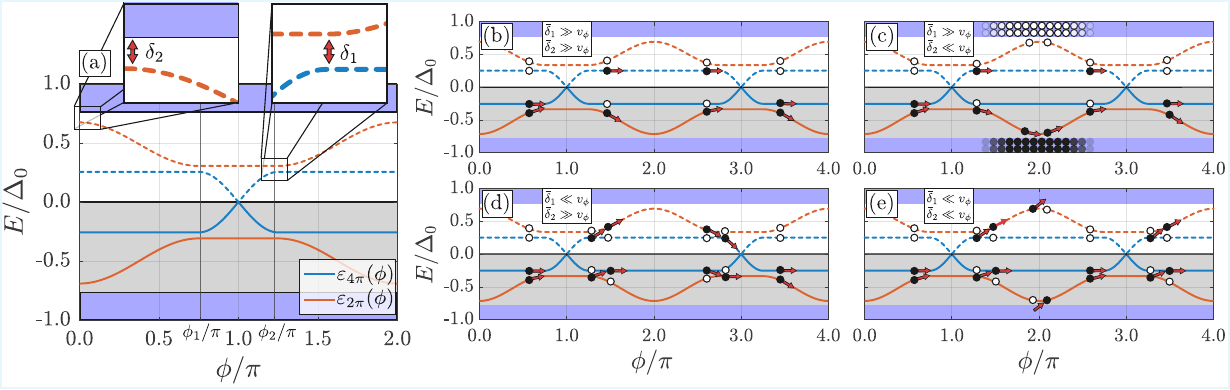}
  \caption{Panel (a): Schematic ABS spectrum that we use to approximate the full tight-binding spectrum shown in Fig.~\ref{fig:setup}(b). The solid (dashed) line represents the states below (above) zero energy. The two insets show the gap $\delta_{1}$ between the $4\pi$- and $2\pi$-periodic ABS and the gap $\delta_{2}$ between the $2\pi$-periodic ABS and the quasiparticle continuum. Panel (b)-(e): The phase dynamics of a single edge for different values of the spectral gaps $\bar\delta_{1,2}$ compared to the phase velocities $v_\phi$ of the particles. Landau-Zener transitions do not occur for $\bar\delta_{1/2}\ll v_\phi$. In panel (c) we schematically represent the occupied quasiparticle continuum  below zero energy and the unoccupied quasiparticle continuum above zero energy.}
  \label{fig:ABS_RSJ}
\end{figure*}

\section{Signatures of the backscattering process}

We shift our focus to studying signatures of the coupling between phase-dependent and phase-independent ABS. First, we use the ABS spectrum obtained in the previous section as input for a phenomenological RSJ model to analyze the Shapiro response. Second, we examine how the same backscattering mechanism affects the SQI pattern.

\subsection{Shapiro steps}

The presence of avoided level crossings energetically isolates the lowest 4$\pi$-periodic ABS from the rest of the 2$\pi$-periodic spectrum. The presence of this gap dynamically suppresses transitions from the ABS into the quasicontinuum under sufficiently adiabatic driving~\cite{San-Jose2012a, Pikulin2012a, Dominguez2012a,Virtanen2013a}. As a result, particles can traverse the full $4\pi$-periodicity of the ABS, without being absorbed into the quasicontinuum at $\varphi=2\pi n$, $n\in\mathbb{Z}$, leading to a suppression of odd-integer Shapiro steps. To capture this mechanism phenomenologically, we employ the resistive shunted junction (RSJ) model adapted for the present scenario, i.e.~two parallel Josephson junctions accounting for the top and bottom QSHI edges ~\cite{Lee2014a,Dominguez2012a,Dominguez2017a}. Each edge hosts a $4\pi$- and a $2\pi$-periodic ABS, $\varepsilon_{4\pi}(\phi)$ and $\varepsilon_{2\pi}(\phi)$, which we approximate based on the full tight-binding calculation, as shown in Fig.~\ref{fig:ABS_RSJ}. Also taking into account the occupation of the ABS yields the following supercurrents
\begin{align}
\label{4piCurrent}
    I_{4\pi}(\phi)&=-(2e/\hbar)\left[(n_{4\pi}^{\text{top}}-1/2)+(n_{4\pi}^{\text{bot}}-1/2)\right]\partial_{\phi} \varepsilon_{4\pi}(\phi),\\
    I_{2\pi}(\phi)&=-(2e/\hbar)\left[(n_{2\pi}^{\text{top}}-1/2)+(n_{2\pi}^{\text{bot}}-1/2)\right]\partial_{\phi}\varepsilon_{2\pi}(\phi),
\end{align}
with $n_{4\pi}^\text{top/bot}$ and $n_{2\pi}^\text{top/bot}$ the occupations of the ABS \textit{below zero energy} for the top (top) and bottom (bot) edge of the QSHI, respectively, see also Fig.~\ref{fig:ABS_RSJ}(b)-(e). In this scenario, the RSJ model takes the following form

\begin{align}
\label{RSJ equation}
I_{\mathrm{dc}} + I_{\mathrm{ac}}\sin(\omega_{\mathrm{ac}} t)
= \frac{\hbar}{2eR}\frac{d\phi}{dt}+I_{2\pi}(\phi) + I_{4\pi}(\phi),
\end{align}
with $I_{\text{dc}}$ and $I_{\text{ac}}=0.2\times I_{4\pi}$ the DC and AC components of the drive, $I_{4\pi}=eE_0/2\hbar$, $\omega_{\text{ac}}= 2eRI_{4\pi}/\hbar=2.2\,$GHz the AC frequency and $R=40.1\,$$\Omega$ the normal state resistance, see App.~\ref{App.RSJ} for more details.

\begin{figure*}[t!]
  \centering
  \includegraphics[width=1\linewidth]{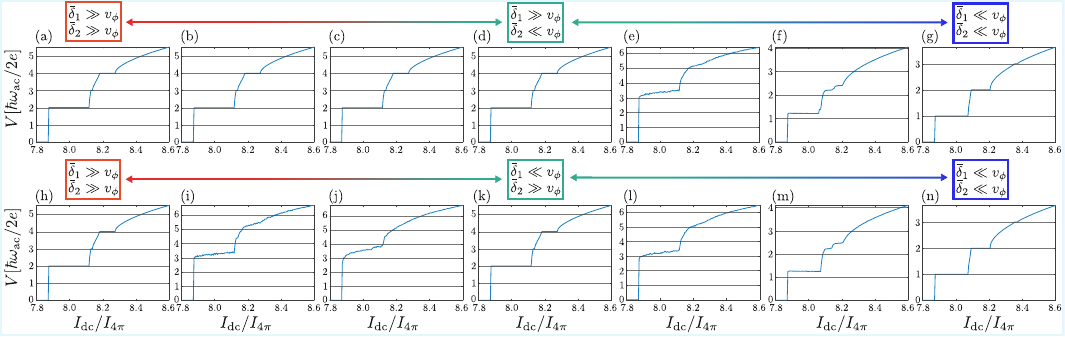}
  \caption{DC Voltage $V$ as a function of the applied $I_{\mathrm{dc}}$ for various values of the Landau--Zener gaps $\bar\delta_{1,2}$ relative to the phase velocity $v_\phi$. 
First row: $\bar\delta_1\gg v_\phi$ remains fixed while $\bar\delta_2$ varies from $\bar\delta_2\gg v_\phi$ in panel~(a) to $\bar\delta_2\ll v_\phi$ in panel~(d). Then also $\bar\delta_1$ decreases from $\bar\delta_1\gg v_\phi$ in panel~(d) to $\bar\delta_1\ll v_\phi$ in panel~(g).
Second row: $\bar\delta_2\gg v_\phi$ remains fixed and $\bar\delta_1$ varies from $\bar\delta_1\gg v_\phi$ in panel~(h) to $\bar\delta_1\ll v_\phi$ in panel~(k). Then also $\bar\delta_2$ decreases from $\bar\delta_2\gg v_\phi$  in panel~(k) to $\bar\delta_2\ll v_\phi$ in panel~(g).}
  \label{fig:VoltageCurves}
\end{figure*}

To account for the effect of the avoided level crossing, we incorporate two distinct non-adiabatic Landau-Zener transitions into the RSJ model: (i) transitions between $\varepsilon_{4\pi}(\phi)$ and $\varepsilon_{2\pi}(\phi)$, shown in Fig.~\ref{fig:ABS_RSJ}(d) and (e), and (ii) transitions between $\varepsilon_{2\pi}(\phi)$ and the quasiparticle continuum, depicted in Fig.~\ref{fig:ABS_RSJ}(e). Based on the locations of the avoided level crossings in the ABS spectrum, we consider LZ transitions between the two ABS at $\phi_1\approx (2n+0.75) \pi$ and $\phi_2\approx (2n+1.25) \pi$ and LZ transitions between the $\varepsilon_{2\pi}(\phi)$ and the quasiparticle continuum at $\phi_c\approx2\pi n$, with $n\in\mathbb{Z}$, as shown in Fig.~\ref{fig:ABS_RSJ}(a). The standard Landau-Zener probability then takes the form~\cite{Dominguez2012a}
\begin{align}
    P^\text{LZ}_{1/2}=\exp\left(-2\pi\frac{ \bar\delta_{1/2}^2}{v_\phi}\right)
\end{align}
with $v_{\phi}= d\phi/d\tau$ the dimensionless phase velocity, $\tau=(2eRI_{4\pi}/\hbar)t$ the dimensionless time, $\bar\delta_{1/2}=\delta_{1/2}/\sqrt{e I_{4\pi} R \Delta_{4\pi/2\pi}}$ the dimensionless Landau-Zener gap and $\delta_{1}=\delta_2\approx 0.1 \Delta_0$ the gaps of the avoided level crossings, represented in Fig.~\ref{fig:ABS_RSJ}(a), see further details in App.~\ref{App.RSJ}.

For $\bar\delta_{1}, \bar\delta_{2}\gg v_{\phi}$ all Landau-Zener transitions are suppressed, i.e.~$P_{1/2}^\text{LZ}\approx 0$. As a result, the occupations of the ABS remain fixed throughout the phase evolution~\footnote{To avoid confusion, we emphasize that the parity configuration, changes whenever the $4\pi$-periodic ABS crosses zero energy, since we define the parity only for negative energies. However, this does not correspond to a change in the actual occupation of the ABS, which remains fixed in this scenario.}, and therefore, the $4\pi$-periodic supercurrent contribution becomes visible in the Shapiro experiment with only even Shapiro steps, see Fig.~\ref{fig:VoltageCurves}(a,h).

Further, by decreasing the ratio $\bar\delta_2/v_\phi$, while keeping $\bar\delta_1\gg v_\phi$, the $4\pi$-periodic ABS remains effectively separated from the quasiparticle continuum by the gap $\bar{\delta}_1$ since non-adiabatic transitions into the quasiparticle continuum are Pauli blocked, as illustrated in Fig.~\ref{fig:ABS_RSJ}(c). In this situation, only even Shapiro steps appear, see Fig.~\ref{fig:VoltageCurves}(a-d). 
Similarly, in the limit of $\bar\delta_2\gg v_\phi$ and $\bar\delta_1\ll v_\phi$, we also observe the suppression of odd-integer Shapiro steps, see Fig.~\ref{fig:ABS_RSJ}(c) and~(d).
In the opposite limit, $\bar\delta_1,\bar\delta_2\ll v_\phi$, the $4\pi$-periodic ABS emits particles to the quasicontinuum at $\phi=2\pi n$, $n\in\mathbb{Z}$, resulting in an effective $2\pi$-periodic ABS, with Shapiro steps at all integer multiples, see Fig.~\ref{fig:VoltageCurves}(f, g, m, n).

In the intermediate regime, where either $\bar{\delta}_1\sim v_\phi$ or $\bar{\delta}_2\sim v_\phi$, $\varepsilon_{4\pi}(\phi)$ is not isolated from the rest of the spectrum, allowing for dynamical transitions between different parities. The resulting Shapiro steps appear at non-quantized values and develop a finite slope, see Fig.~\ref{fig:VoltageCurves}~(e, f, i, j, l, m). This occurs because the parity switches between two different configurations, one where the supercurrents add up, and one where the supercurrents cancel, e.g.~$(1,1,1,1)$ and $(1,0,1,0)$, respectively. In this situation, the total voltage is approximately given by weighted average of the voltage developed on each parity configuration, see more details in App.~\ref{Parity configuration} and \ref{App: Dwell-Time weighted average}.

\subsection{SQI pattern}

Another way to probe the backscattering mechanism is through the SQI pattern. Since the backscattering introduced here strongly modifies the spatial distribution of the supercurrent, it also leaves clear signatures in the critical current $I_\text{c}$ as a function of the external magnetic flux $\Phi = BWL_N$, also known as the SQI pattern~\cite{Dynes1971a}.
The change in the SQI pattern is present for parity constrained or unconstrained scenarios. Thus, we calculate $I_\text{c}$ from the maximum of the equilibrium supercurrent $I_\text{c} =\text{Max}_\phi\{I(\phi)\}$, with 
\begin{align}
\label{eq:statcurr}
I(\phi)= \frac{e}{\hbar}  \int dE ~\text{Tr}_W\left\{[V_{LR} G^{+-}_{RL}(E)-V_{RL} G^{+-}_{LR}(E)]_e \right\},
\end{align}
where $G_{RL}^{+-}(E)\equiv G^{+-}(x_0-a,x_0+a,E)$ are the equilibrium lesser GFs evaluated at $L=x_0-a_x$ and $R=x_0+a_x$ positions, with $x_0$ placed on the central normal part of the junction. Besides, $V_{LR/RL}$ couples different layers of the discretized Hamiltonian. Here, we take the trace over the electron part of the GFs and the width $\text{W}$ of the junction.

\begin{figure}[tb]
  \centering
  \includegraphics[width=1\linewidth]{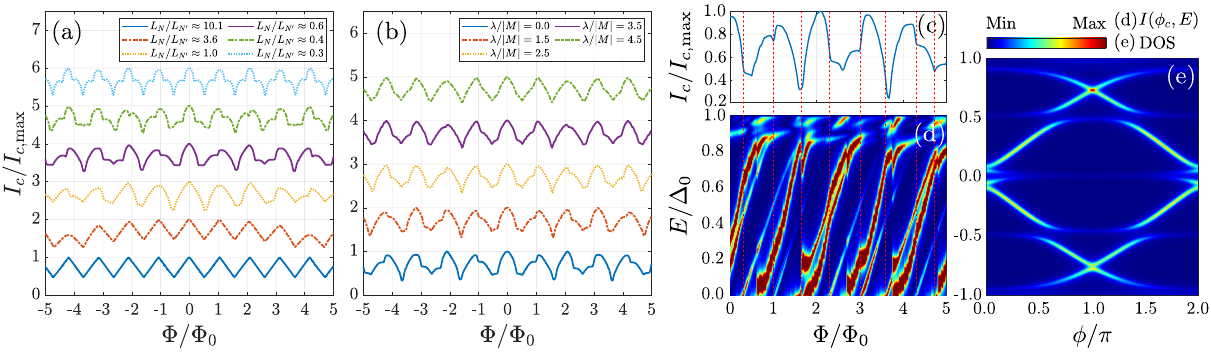}
  \caption{The SQI pattern for Josephson junctions with the same parameters as in Fig.~\ref{fig:setup}(b) but for (a) different $L_{N'}/a=7, 20,70,120,170,220$ and (b) in the presence of disorder with increasing disorder strengths $\lambda$ for $L_{N'}/a=100$. For better visibility, the SQI curves are shifted upwards by a constant $\Delta I_\text{c}=I_{c,max}$.
  Panel~(c): critical current vs magnetic flux. Panel~(d): integrand of Eq.~\eqref{eq:statcurr} as a function of energy and $\Phi/\Phi_0$, for $\phi=\phi_\text{c}$ fulfilling $I_\text{c}=I(\phi_\text{c})$, with the same parameters as in Fig.~\ref{fig:setup}(b). Panel~(e): Andreev bound spectrum as a function of $\phi$, with $\Phi$ fulfilling the condition~\eqref{eq.condition}, where the phase independent ABS hybridizes at zero energy with the phase-dependent ABS. We set $L_s=0.33\,\mu$m to make the gaps more visible. }
  \label{fig:IcvsEvsflux}
\end{figure}%\noindent
 
In Fig.~\ref{fig:IcvsEvsflux}(a), we study the impact of considering different ratios $L_{N'}/L_N$ on the SQI pattern. 
Thus, we increase $L_{N'}$ in steps of $50\,a$ and show the normalized $I_\text{c}/I_{\text{c,max}}$ for an increasing $L_{N'}$ from bottom to top. Here, $I_{\text{c,max}}$ is the maximum value of $I_\text{c}$ over $\Phi$ for a given set of parameters. For clarity, we shift vertically each curve as we increase $L_{N'}$. We can observe that deviations of the sinusoidal character of the SQI pattern arise for $\Phi/\Phi_0\sim L_N/ L_{N'}$, e.g.~for $L_N/ L_{N'}\approx 3.6$, we observe changes for $\Phi/\Phi_0 \geq 3$, as the area enclosed by the N' regions needs a comparable magnetic flux. For larger $L_{N'}$, the SQI pattern changes its periodicity for smaller magnetic flux, developing erratic patterns or periodic ones, depending on the relative fluxes threading the N and N' regions. Note that introducing different lengths for the left and right $L_{N'}$ can lead to an even more erratic pattern, since in this situation, there is an extra periodicity coming into play, as discussed in more detail in App.~\ref{App.SQI}.

The presence of potential disorder on the N' regions broadens the energy spectrum of the phase-independent ABS, as shown in App.~\ref{App.disorder}. Consequently, the coupling between phase-dependent and -independent ABS becomes effectively suppressed. To see its influence on the SQI pattern, we introduce a random local potential on the sites placed at the N' regions within the values $[-\lambda,\lambda]$. In Fig.~\ref{fig:IcvsEvsflux}(b), we can observe that the features distorting the SQI pattern are smoothed out for increasing $\lambda$, resulting from the broadening of the phase-independent ABS. Indeed, for $\lambda\lesssim 3|M|$, we observe numerically that the ABS become broader, but nevertheless, well-formed independently of $p_{N'}$, see App.~\ref{App.disorder}. For larger disorder strengths $\lambda> 4|M|$, the topological protection of the QSH edges is effectively removed since QSH edge states can now hybridize with the bulk states.
Consequently, the phase-independent ABS do not form, yielding approximately the conventional picture of QSH Josephson junction~\cite{Baxevanis2015a}. 

To have a direct proof of the phase-independent ABS participation, we propose to selectively remove the topological character by means of a finite magnetic flux. Due to the different areas in the N and N' regions, a finite magnetic flux shifts relatively the position of phase-independent and phase-dependent ABS. Indeed, the former become 
\begin{align}
\label{eq:analyticalresult}
E_n=\frac{\hbar v_F}{p_{N'}}\pi (n+1/2+\alpha \Phi/\Phi_0),
\end{align}
with $n\in \mathbb{Z}$ and $\alpha=L_{N'}/L_N$. 
Hence, setting $E_n\approx 0$ at the maximum of the SQI, namely, 
\begin{align}
\Phi/\Phi_0\approx -\text{int}\{(n+1/2)/\alpha\},\label{eq.condition}
\end{align} 
we can gap out the MBS, removing the fractional Josephson effect. This change can be probed by means of the Shapiro experiment or the Josephson radiation. 

To see this phenomenon explicitly, we represent in Fig.~\ref{fig:IcvsEvsflux}(c) the SQI pattern together with the integrand of Eq.~\eqref{eq:statcurr}, henceforth called $I(\phi_c,E)$, as a function of the energy $E$ and $\Phi/\Phi_0$ in panel~(b)~\footnote{Note that for each value of $\Phi$, we use the phase difference $\phi_\text{c}$, fulfilling $I_\text{c}(\Phi)=I(\phi_\text{c},\Phi)$.}. Here, we can observe two types of (roughly) periodic contributions, one with an $\int$-shape and period $\Phi/\Phi_0\approx 1$, which results from the ABS localized at the SNS region, and a linear dispersion, with period $\alpha \Phi/\Phi_0\approx 1$, which arises from the phase-independent ABS. Remarkably, the largest distortion on the SQI pattern~(a) coincides with the presence of phase-independent ABS at zero energy, see vertical red dashed lines. Moreover, when both contributions coincide in energy and flux, they develop a minimum resulting from the hybridization process between them. 
As we have anticipated, when both contributions coincide around zero energy around an integer value of $\Phi/\Phi_0$, the topological state turns into a trivial one. We show the resulting ABS as a function of $\phi$ for $\Phi/\Phi_0=1$ in Fig.~\ref{fig:IcvsEvsflux}(d). We observe a gap opening around zero energy, yielding a 2$\pi$-periodic ABS. Note that the lack of particle-hole symmetry in the spectrum originates from the combined reduction of the Nambu basis and the presence of a finite magnetic flux. We recover the particle-hole symmetry when repeating the calculation with the complete basis, see App.~\ref{App.SQI}.

\section{Conclusion}
 Using a non-interacting tight-binding model to describe an extended quantum spin-Hall Josephson junction N'SNSN' under time-reversal symmetry, we design a backscattering mechanism that isolates energetically a $4\pi$-periodic contribution from the rest of the spectrum. In this geometry, additionally to the phase-dependent Andreev bound states localized at the central SNS junction, extra phase-independent Andreev bound states form at the edges of the N'S regions with discrete energy states $E_n=\pi \hbar v_F(n+1/2)/p_{N'}$ determined by the perimeter of the N' part $p_{N'}$. Thus, in a scenario with transparent enough superconducting leads, these additional ABS mediate a backscattering mechanism between opposite edges on the SNS junction, resulting in the appearance of avoided level crossings when both types of Andreev bound states are at resonance, yielding a measurable $4\pi$-periodic gap $\Delta_{4\pi,\text{eff}}= E_0=\pi \hbar v_F/2p_{N'}\sim 0.57-0.11\,$meV, for $p_{N'}\sim 1-5\,\mu$m. The resulting ABS remains topological~\cite{Crepin2014a,Lee2014a} and can be probed, for example, by means of the Shapiro experiment or the Josephson radiation if the driving is adiabatic enough~\cite{San-Jose2012a, Pikulin2012a, Dominguez2012a, Virtanen2013a}. To this end, we analyze the effective periodicity using a phenomenological RSJ model that incorporates parity switching between the $4\pi$- and $2\pi$-periodic ABS and between the $2\pi$-periodic ABS and the quasiparticle continuum. Using the avoided level crossings obtained from the tight-binding calculations as the Landau-Zener gaps, we find regimes in which the odd Shapiro steps disappear, providing a direct signature of the isolated $4\pi$-periodic ABS.

An important question in this context is whether quasiparticle poisoning can disrupt the effective $4\pi$-periodicity. Estimates for the quasiparticle poisoning rates range from $\tau_{qp}=10\,$ns -- $10\,$$\mu$s,  depending on the microscopic details~\cite{Rainis2012}. In our RSJ description, the relevant time scale is set by the AC driving frequency $\tau_{\text{ac}}=1/2\pi\omega_{\text{ac}}=2.8\,$ns. Since $\tau_{\text{ac}}$ is comparable or shorter than the estimated poisoning times, quasiparticle poisoning does not necessarily obstruct the observation of the $4\pi$-periodic signal.

Furthermore, we have tested the stability of the phase-independent ABS against the presence of potential disorder finding a robust behavior for disorder strengths larger than the topological gap, i.e.~$\lambda\lesssim 3|M|$ and can extend on the order of several microns. 

We predict signatures of this backscattering mechanism to be present in the SQI pattern, which becomes distorted due to the additional trajectories enclosing the N'S regions. These trajectories gather an additional magnetic flux and hence, introduce a new periodicity into the sinusoidal $\Phi_0$-periodic SQI pattern, which develops local maxima and minima on the magnetic flux scale of $\sim(L_{N}/L_{N'})\Phi_0$. Moreover, due to the difference between $L_N$ and $L_{N'}$, we can use the magnetic flux to tune phase-independent Andreev bound states towards the phase-dependent ones. We find that when these two types of ABS become close to zero energy, we remove the 4$\pi$-periodicity selectively and, therefore, it can be used as a control knob to switch on and off the fractional Josephson effect.

In conclusion, we believe that the use of this backscattering mechanism can contribute decisively to the design of topological Josephson junctions with 4$\pi$-periodic ABS energetically isolated from the quasicontinuum under time-reversal symmetry as it naturally uses the topology and associated robustness of the QSH effect. We analyzed in detail two experimental probes of the proposed effect, the Shapiro steps and the SQI pattern. Moreover, the physics described here is not restricted to the given N'SNSN geometry. Indeed, analogous ideas can be extended onto a single edge of two topological Josephson junctions in series, i.e.~the SNSN'S junction. Here, the central N' region develops phase-independent ABS when the two inner (outer) superconductors share the same phase difference.

\section*{Acknowledgements}
We acknowledge stimulating discussions with B. Trauzettel, E. Bocquillon, E. M. Hankiewicz, N. Traverso Ziani, M. Stehno and L. W. Molenkamp. F.D.\ and P.R.\ gratefully acknowledge funding by the Deutsche Forschungsgemeinschaft (DFG, German Research Foundation) under Germany’s Excellence Strategy – EXC-2123 QuantumFrontiers – 390837967.

\begin{appendix}

\section{ ~~Details on the Transport Formalism}
\label{App.transport}
We are interested in modelling a discrete $2D$ material, therefore, we apply the tight-binding discretization method on  the Hamiltonian in Eq.~\eqref{eq:BdGBHZ} by replacing the continuous momentum operators $\hat{k}_{x,y}=-i \partial_{x,y}$ by their discretized versions in the Hamiltonian of Eq.~(\ref{eq:BdGBHZ})\cite{Datta1995a}:
\begin{align}
    \partial_{x/y}\Psi(x,y)&\approx \frac{1}{2a_{x/y}}\left( \Psi_{j_{x/y}+a_{x/y}}-\Psi_{j_{x/y}-a_{x/y}} \right)\\
    \partial^2_{x/y}\Psi(x,y)&\approx \frac{1}{a^2_{x/y}}\left( \Psi_{j_{x/y}+a_{x/y}}-2\Psi_{j_{x/y}}+\Psi_{j_{x/y}-a_{x/y}} \right),
\end{align}
where $a_{x,y}$ are the lattice constants of the two-dimensional lattice and $j_{x,y}$ is the site index in $x$ and $y$ direction.
Until specified otherwise, we will use $a=a_{x}=a_{y}=5\,$nm as the lattice constant. The corresponding tight-binding Hamiltonian thus takes the following form
\begin{align}
\label{TBHam}
    \mathcal{H}=\sum_{n=1}^{L} \sum_{m=1}^{W} \mathcal{H}_{\substack{nn \\ mm}}^{0}+ \mathcal{V}_{\substack{n,n+1 \\ m,m}}+\mathcal{V}_{\substack{n,n \\ m,m+1}}+ \mathcal{V}_{\substack{n,n+1 \\ m,m}}^{\dagger}+\mathcal{V}_{\substack{n,n \\ m,m+1}}^{\dagger},
\end{align}
where L and W represent the length and width of the junction and the indices $n$ and $m$ represent the spatial dimensions $x$ and $y$, respectively.
$\mathcal{H}^{0}$ is the on-site Hamiltonian and $\mathcal{V}$ describes the hoping between neighbouring sites. 

To efficiently treat the tight-binding Hamiltonian of Eq.~\eqref{TBHam}, we employ a standard recursive Green's method only for the $x$-direction of the Hamiltonian, i.e. we do not perform the sum over $L$ but instead grow the lattice recursively with Green's functions. To see what this means, we first rewrite Eq.~\eqref{TBHam} as
\begin{align}
    \mathcal{H}&=\sum_{n=1}^{L} H_{nn}^{0}+ V_{n,n+1}+V_{n,n+1}^{\dagger}, \\
    H_{nn}^{0}&=\sum_{m=1}^W \mathcal{H}_{\substack{nn \\ mm}}^{0}+\mathcal{V}_{\substack{n,n\\m,m+1}}+\mathcal{V}_{\substack{n,n\\m,m+1}}^{\dagger}, \\
    V_{n,n+1}&=\sum_{m=1}^W \mathcal{V}_{\substack{n,n+1 \\ m,m}},
\end{align}
where $H_{nn}^{0}$ describes the on-site Hamiltonian of the $n$-th stripe of $\text{dim}(H_{nn}^{0})=4\bar{W}\times 4\bar{W}$ and $V_{n,n+1}$ describes the hopping from the $n$-th stripe to the $n+1$-th stripe. 
Since the system has a homogeneous hopping in the $x$-direction, we simplify the notation $V_{LR}\equiv V_{n,n+1}$ and $V_{RL}\equiv V_{n,n+1}^{\dagger}$.
Using this notation, we find the perturbed retarded/advanced Green's function $G_n^{r/a}$ by the recursive scheme~\cite{MacKinnon1985}
\begin{align}
    G_{nn}&=\left[ g^{-1}_{nn}-V_{RL}G_{n-1,n-1}^{L}V_{LR}-V_{LR}G_{n+1,n+1}^{R}V_{RL}\right]^{-1},\\
    \label{eq:LeftSFGF}
    G^{L}_{nn}&=\left[ g^{-1}_{nn} -V_{RL}G^{L}_{n-1,n-1}V_{LR}\right]^{-1}, \\
    \label{eq:RightSFGF}
    G^{R}_{nn}&=\left[ g^{-1}_{nn} -V_{LR}G^{R}_{n+1,n+1}V_{RL}\right]^{-1},
\end{align}
where we lightened the notation by omitting the superscript $r/a$. $g_{nn}^{r/a}=[(E\pm i\eta)\mathbb{1}_{4 \bar{W} \times 4 \bar{W}} -H_{nn}^{0}]^{-1}$ is the unperturbed Green's function for the $n$-th stripe. Furthermore, we set $G_{00}^{R}=G_{00}^{L}=\mathbb{0}$, with $\mathbb{0}$ as the zero-matrix, since in the first iteration we only have a single site, which is by definition unperturbed $G_{11}^{R}=G_{11}^{L}=g_{11}$.
We obtain the DOS by taking the imaginary part of the perturbed advanced Green's function 
\begin{align}
    \text{DOS}(E)=\frac{1}{\pi} \text{Im} \sum_{n=1}^{N}\text{Tr}_{W}\{G_{nn}^{a}\},
\end{align}
where $\text{Tr}_{W}$ is the trace with respect to the width W and $N$ is the number of sites we calculate the DOS on and which can be chosen to be smaller than $L$. 
\begin{figure}[tb]
  \centering
  \includegraphics[width=1\linewidth]{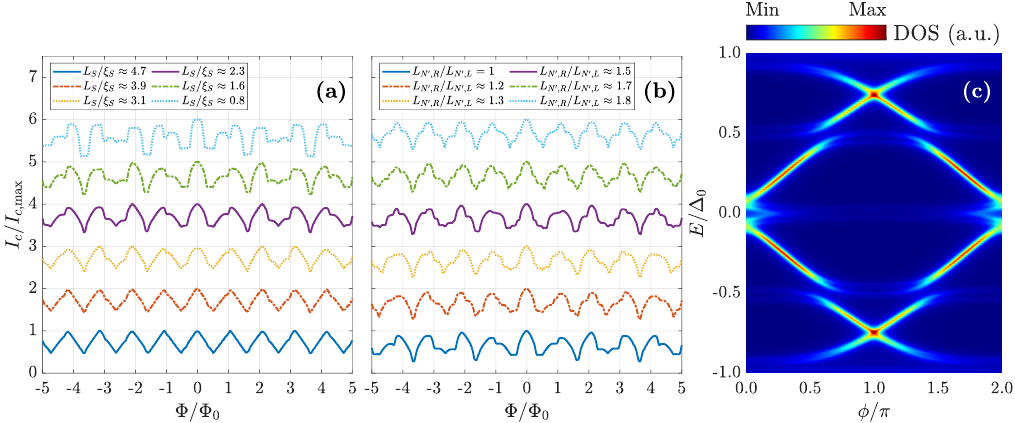}
  \caption{Panel~(a): critical current versus magnetic flux for (a) different lengths of the superconductor $L_S/a=180, 150, 120, 90, 60, 30$ and (b) different lengths of the external parts. We fix $L_{N’,L}/a=120$ and increase the length of the right part to $L_{N',R}/a=140,160,180,200,220$. For visibility, the SQI curves are shifted upwards a constant $\Delta I_\text{c}=I_{c,max}$. Panel (c): The Andreev bound state spectrum for the full BHZ Hamiltonian with both spin degrees of freedom.}
  \label{fig:SQISM}
\end{figure}\noindent

To obtain the supercurrent of the Josephson junction, we imagine splitting the junction along the $x$-direction in a left $(L)$ and a right $(R)$ part. The current then takes the following form
\begin{align}
\label{eq:statcurr2}
I(\phi)=  \frac{e}{\hbar}  \int_{-\infty}^{+\infty} dE \,\ \text{Tr}_W\left\{[V_{LR} G^{+-}_{RL}(E)-V_{RL} G^{+-}_{LR}(E)]_e \right\},
\end{align}
where, in equilibrium, the lesser Green's functions $G^{+-}$ simplify to
\begin{align}
\label{eq:LeftNoneq}
G^{+-}_{LR}(E)&=\left[ G^a_{LR}(E)-G^r_{LR}(E)\right] f(E), \\
\label{eq:RightNoneq}
G^{+-}_{RL}(E)&=\left[ G^a_{RL}(E)-G^r_{RL}(E)\right] f(E),
\end{align}
with $f(E)$ as the equilibrium Fermi-Dirac distribution.
To make use of Eq.~\eqref{eq:statcurr2} we relate the Green's functions of Eq.~\eqref{eq:LeftNoneq} and (\ref{eq:RightNoneq}) to the surface Green's functions of Eq.~\eqref{eq:LeftSFGF} and (\ref{eq:RightSFGF}) as follows
\begin{align}
&G_{LR}^{r/a} =\tilde{g}_{LL}^{r/a} V_{LR} G_{RR}^{r/a},\\ 
&G_{RL}^{r/a} =G_{RR}^{r/a} V_{RL} \tilde{g}_{LL}^{r/a}\text{~~~and}\\
&G_{RR}^{r/a}=[(\tilde{g}^{r/a}_{RR})^{-1}-V_{RL}\mathcal{G}^{r/a}_{LL}V_{LR}]^{-1},
\end{align}
where $\tilde{g}_{LL/RR}^{r/a}=G_{nn}^{L/R}$ represents the retarded/advanced surface Green's function of the left/right side, given in Eqs.~(\ref{eq:LeftSFGF}) and (\ref{eq:RightSFGF}). Furthermore, we made use of the discretized form of the Dyson equation
\begin{align}
    &\langle i|G|j \rangle=\langle i|G+\mathcal{G}VG|j \rangle\\
 \Leftrightarrow \quad & \mathcal{G}_{ij}=G_{ii}\delta_{ij}+\mathcal{G}_{ii}V_{i,i-1}G_{i-1,i}+\mathcal{G}_{ii}V_{i,i+1}G_{i+1,j}. 
\end{align}

\section{Additional SQI calculations}
\label{App.SQI}
In addition to Fig.~\ref{fig:IcvsEvsflux} (a) and (b), we investigate the SQI pattern resulting from varying $L_{s}$, which controls the coupling strength between the phase-dependent and independent ABS, depicted in Fig.~\ref{fig:SQISM}(a). We set $L_{N'}=0.6\,\mu$m and reduce $L_s$ in steps of $0.15\,\mu$m from $L_s=0.9\,\mu$m to $L_s=0.1\,\mu$m. For clarity, we have shifted the curves upwards for a decreasing $L_s$. For $\xi_s/L_s \ll 1$, we observe that the SQI pattern exhibits the typical sinusoidal pattern with periodicity $\Phi_0$, denoting the lack of coupling to the N' parts. As we decrease $L_s$, the superconducting leads become more transparent, enhancing the effective coupling between opposite edges. The resulting SQI pattern develops local maxima and minima on top of the sinusoidal pattern, and turns into an erratic pattern for $\xi_s/L_s > 1$. Moreover, we note that the non-normalized $I_\text{c}$ becomes more suppressed by reducing $L_s$ (not shown), since the avoided level crossings become more dominant, resulting in a flatter ABS spectrum.

In Fig.~\ref{fig:SQISM}(b) we consider the scenario of a different left and right $L_{N'}$ parts, which introduces additional periodicities relative to panel~\ref{fig:IcvsEvsflux}(b). Due to the period mismatch of the left and right external parts, the SQI exhibits even more erratic patterns, compared to Fig.~\ref{App.SQI} and Fig.~\ref{fig:SQISM}(a).
In general, it is therefore difficult to predict a given periodicity. This is in contrast to the SQI patterns resulting from the direct (energy-independent) coupling between opposite edges, which can lead to an even-odd effect~\cite{Baxevanis2015a, Tkachov2015a, Haidekker2020a, Vigliotti_2022, Viglioti2023a, Dominguez2024a}.

\begin{figure}[t]
  \centering
  \includegraphics[width=6.3in]{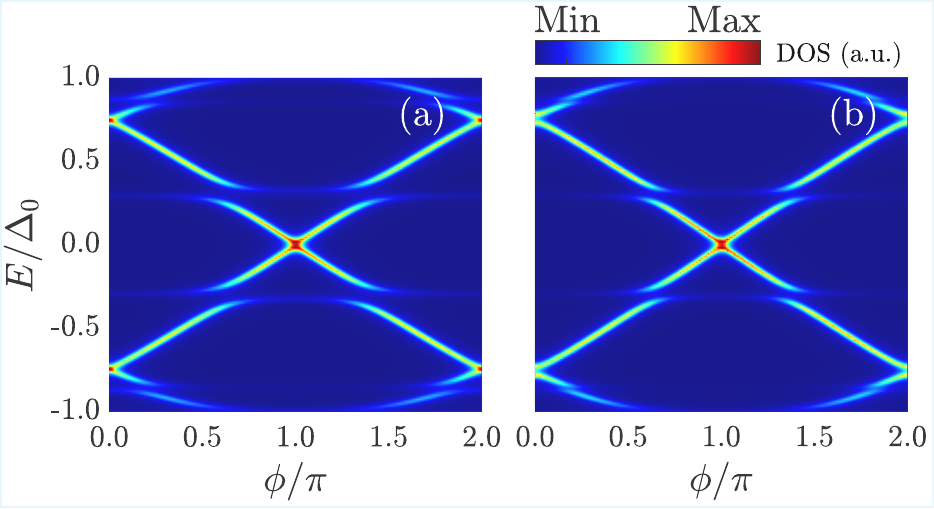}
  \caption{The ABS for a (a) N'SNSN' Josephson junction and a (b) N'SNS Josephson junction.}
  \label{fig:OneTwoQSHI}
\end{figure}

In the main text, we consider a single spin-degree of freedom of the original BHZ Hamiltonian. As a consequence, the system is not particle-hole symmetric in the presence of a magnetic flux, as shown in Fig.~\ref{fig:IcvsEvsflux}(e). However, we can restore the particle-hole symmetry upon including the second spin degree of freedom, as shown in Fig.~\ref{fig:SQISM}(c).

\section{ ~~N'SNS -junction}
\label{App: N'SNS -junction}
In this section, we demonstrate that the backscattering mechanism also works with only one external QSHI. For this we omit the right N' part of the Josephson junction presented in the main text and compute the LDOS. Comparing the N'SNSN' junction of the main text in  Fig.~\ref{fig:OneTwoQSHI} (a) to the N'SNS junction in  Fig.~\ref{fig:OneTwoQSHI} (b) we see that they yield a similar avoided level crossing.

\section{ ~~RSJ model}
In this section, we explain in more detail how we set up the RSJ model used in the main text to calculate the Shapiro steps. We also explain the role of the ABS parity, and we provide an alternative explanation for the appearance of Shapiro steps at non-integer values in regimes of intermediate Landau–Zener transitions.
\label{App.RSJ}
\subsection{Andreev bound states and RSJ equation}
\label{}
In the following, we describe in more detail the resistively shunted junction (RSJ) model used to compute the Shapiro response presented in the main text. We start by constructing the supercurrents from the ABS, which we obtained from the full tight-binding calculation, shown in Fig.~\ref{fig:setup}(b).
To approximate the ABS spectrum, we define the $4\pi$-periodic ABS $\varepsilon_{4\pi}(\phi)$ and the $2\pi$-periodic ABS $\varepsilon_{2\pi}(\phi)$ piecewise as functions of the superconducting phase difference $\phi$ as
\begin{align}
\varepsilon_{4\pi}(\phi)&=\Delta_{4\pi}\times
\begin{cases}
1 & 0 \le \phi < \phi_1,\\[6pt]
\cos\!\left(\dfrac{\phi-\phi_1}{\phi_2-\phi_1}\pi\right), & \phi_1 \le \phi < \phi_2,\\[6pt]
-1, & \phi_2 \le \phi < 2\pi+\phi_1,\\[6pt]
-\cos\!\left(\dfrac{\phi-(2\pi+\phi_1)}{\phi_2-\phi_1}\pi\right), & 2\pi+\phi_1 \le \phi < 2\pi+\phi_2,\\[6pt]
1, & 2\pi+\phi_2 \le \phi \le 4\pi
\end{cases}
\qquad (\text{defined mod } 4\pi)
\\[1em]
\varepsilon_{2\pi}(\phi)
&=\Delta_{4\pi}+\Delta_{2\pi}\times
\begin{cases}
 \cos^{2}\!\left(\dfrac{\phi}{2\phi_1}\pi\right),
& 0 \le \phi < \phi_1, \\[10pt]
0,
& \phi_1 \le \phi < \phi_2, \\[10pt]
\sin^{2}\!\left(
\dfrac{\phi-\phi_2}{2(2\pi-\phi_2)}\pi
\right),
& \phi_2 \le \phi < 2\pi 
\end{cases}
\qquad (\text{defined mod } 2\pi)
\end{align}
with $\phi_1=0.75\pi$, $\phi_2=1.25\pi$ as the locations of the avoided crossings, shown in Fig.~\ref{fig:ABS_RSJ}(a) and the effective superconducting gaps $\Delta_{4\pi}=E_0$ and $\Delta_{2\pi}=E_1-E_0$ which we also obtain from the tight-binding calculation of the main text. 
$\varepsilon_{4\pi}(\phi)$ is the lowest $4\pi$-periodic ABS and $\varepsilon_{2\pi}(\phi)$ the second lowest $2\pi$-periodic contribution, see Fig.~\ref{fig:ABS_RSJ}(a). The energy of the junction also depends on the occupation of the Andreev bound states, consequently the individual energies take the following form
\begin{align}
    \label{4piABS}
    E_{4\pi}(\phi)&=-\left((n_{4\pi}^{\text{top}}-1/2)+(n_{4\pi}^{\text{bot}}-1/2)\right)\varepsilon_{4\pi}(\phi),\\
    \label{2piABS}
    E_{2\pi}(\phi)&=-\left((n_{2\pi}^{\text{top}}-1/2)+(n_{2\pi}^{\text{bot}}-1/2)\right)\varepsilon_{2\pi}(\phi)
\end{align}
where $n_{4\pi}^{top/bot}$ and $n_{2\pi}^{top/bot}$ describes to the occupation of $\varepsilon_{4\pi}(\phi)$ and $\varepsilon_{2\pi}(\phi)$ \textit{below zero energy} for the top (top) and bottom (bot) edge of the QSHI, respectively. The Bogoliubov–de Gennes Hamiltonian obeys particle–hole symmetry, which ensures that every quasiparticle state at energy $+E$ has a corresponding partner at 
$-E$. The corresponding operators satisfy $\gamma_{E}^{\dagger}=\gamma_{-E}$
, meaning that creating a quasiparticle at $+E$ is equivalent to destroying its partner at $-E$. Consequently, the many-body state is completely described by the occupation of the negative-energy ABS given as $(n_{2\pi}^{\text{bot}},n_{2\pi}^{\text{top}},n_{4\pi}^{\text{bot}},n_{4\pi}^{\text{top}})$. Note that, since we define the parity solely from the negative-energy states, the $4\pi$-periodic ABS switches its occupation at $\phi = (2n+1)\pi$. At these phase values the ABS branch passes through the protected zero-energy crossing, which results in a change of the ground-state parity, as depicted in Fig.~\ref{fig:ABS_RSJ}(a).

We describe the phase dynamics of the junction within the RSJ model, where the supercurrent is the derivative of the Andreev bound states with respect to the superconducting phase difference $I(\phi)=(2e/\hbar)\partial_{\phi}E(\phi)$. Consequently, the ABS of Eqs.~(\ref{4piABS}) and (\ref{2piABS}) yield the following supercurrents
\begin{align}
\label{eq:4piCurrent}
    I_{4\pi}(\phi)&=-4 I_{4\pi}\left((n_{4\pi}^{\text{top}}-1/2)+(n_{4\pi}^{\text{bot}}-1/2)\right)\partial_{\phi} \varepsilon_{4\pi}(\phi)/\Delta_{4\pi},\\
    I_{2\pi}(\phi)&=-4I_{2\pi}\left((n_{2\pi}^{\text{top}}-1/2)+(n_{2\pi}^{\text{bot}}-1/2)\right)\partial_{\phi}\varepsilon_{2\pi}(\phi)/\Delta_{2\pi}
\end{align}
where $I_{\alpha}=e\Delta_{\alpha}/2\hbar$, with $e$ as the electric charge and $\alpha= 2\pi,~4\pi$. In this scenario, the RSJ model takes the form
\begin{align}
\label{RSJ equation App}
I_{\mathrm{dc}} + I_{\mathrm{ac}}\sin(\omega_{\mathrm{ac}} t)
= \frac{\hbar}{2eR}\frac{d\phi}{dt} + I_{\mathrm{tot}}(\phi),
\end{align}
where $I_{\mathrm{dc}}$ and $I_{\mathrm{ac}}$ denote the DC and AC components of the applied current,
$I_{\mathrm{tot}}(\phi)=I_{2\pi}(\phi)+I_{4\pi}(\phi)$ is the total supercurrent,
$\omega_{\mathrm{ac}}$ is the drive frequency, and $R$ is the normal-state resistance.
Due to the size of the effective superconducting gaps $\Delta_{2\pi}$ and $\Delta_{4\pi}$, the two amplitudes of the supercurrents obey $I_{4\pi} < I_{2\pi}$.
Thus, the dominant $2\pi$-periodic contribution can mask the weaker $4\pi$-periodic term, leading to the appearance of Shapiro steps at all integers. However, by choosing $I_{\text{ac}}$ and $\omega_{\text{ac}}$ appropriately, the $4\pi$-periodic contribution can be made visible in the phase dynamics.
To see this, recall that a purely $2\pi$-periodic supercurrent produces a
$2\pi$-periodic washboard potential. Adding a $4\pi$-periodic term modulates this 
potential such that in odd-sectors $4(n-1)\pi \le \phi < 4(n-\tfrac12)\pi$, $n\in \mathbb{Z}$, the 
$4\pi$ component reduces the local slope (slowing the phase), while in even-sectors
$4(n-\tfrac12)\pi \le \phi < 4n\pi$ it increases it (speeding up the phase)~\cite{Dominguez2017a}. By setting $\omega_{\text{ac}}= 2e R I_{4\pi}/\hbar$ and $I_{\text{ac}}=0.2\times I_{4\pi}$ the phase gets trapped in the odd-sectors, but flows freely in the even-sectors. As a consequence, the phase advances in multiples of $4\pi$ per period, which suppresses the odd-integer Shapiro steps in the voltage response.

To solve Eq.~(\ref{RSJ equation App}) numerically, we discretize time into uniform steps and then use a forth-order Runge-Kutta algorithm to solve for the superconducting phase difference $\phi$. From the time-evolution of $\phi$, we then calculate the time-averaged voltage that the junction produces by using the second Josephson equation $V=(\hbar/2e)d\phi/dt$, which yields
\begin{align}
\label{Average voltage RSJ}
    \langle V\rangle =\frac{\phi(T_2)-\phi(T_1)}{(T_2-T_1)\Tilde{\omega}_{ac}}\times \frac{\hbar \omega_{\text{ac}}}{2e},
\end{align}
where $\tilde\omega_{\text{ac}}=(\hbar/2eRI_{4\pi})\omega_{\text{ac}}$ is the dimensionless ac driving frequency.

Finally, we also take into account LZ transitions between $\varepsilon_{4\pi}(\phi)$ and $\varepsilon_{2\pi}(\phi)$ as well as transitions between $\varepsilon_{2\pi}(\phi)$ and the quasiparticle continuum, as described in the main text. The way we incorporate LZ transitions in the RSJ model is by monitoring the time evolution of the phase. Once the phase reaches a potential avoided level crossing, we calculate the Landau-Zener probability with the given instantaneous velocity from the immediate previous time step. Then, we generate a random number $X$ and compare it to $P^\text{LZ}_{1/2}$. If $P^\text{LZ}_{1/2}>X$, we accept the Landau-Zener transition, otherwise we discard it. If a transition occurs between $\varepsilon_{4\pi}(\phi)$ and $\varepsilon_{2\pi}(\phi)$, then the two ABS exchange parity, e.g. $n^{\alpha}_{4\pi}\rightarrow n^{\alpha}_{4\pi}+1$ and $n^{\alpha}_{2\pi}\rightarrow n^{\alpha}_{2\pi}-1$, where $\alpha=top$ or $bot$, see Fig.~\ref{fig:ABS_RSJ}(d). Note here, that this process can only occur when the two ABS are different in parity, i.e. one is empty and the other one is occupied. On the other hand, if a transition occurs between $\varepsilon_{2\pi}(\phi)$ and the quasi-particle continuum then the parity of the ABS is lost to the continuum and we get $n_{2\pi}^\alpha\rightarrow n_{2\pi}^\alpha-1$, as shown in Fig.~\ref{fig:ABS_RSJ}(e).

\subsection{ ~~ Parity configuration}
\label{Parity configuration}

In this subsection, we will discuss in more detail how different initial parity configurations affect the generated voltage. To this end, we initialize the system given in Eq.~(\ref{RSJ equation}) in all $16$ distinct parity configurations labeled by $(n_{2\pi}^{\text{bot}},n_{2\pi}^{\text{top}},n_{4\pi}^{\text{bot}},n_{4\pi}^{\text{top}})$ for different limits of the gaps $\bar\delta_1$ and $\bar\delta_2$.

\begin{figure}[tb]
  \centering
  \includegraphics[width=1\linewidth]{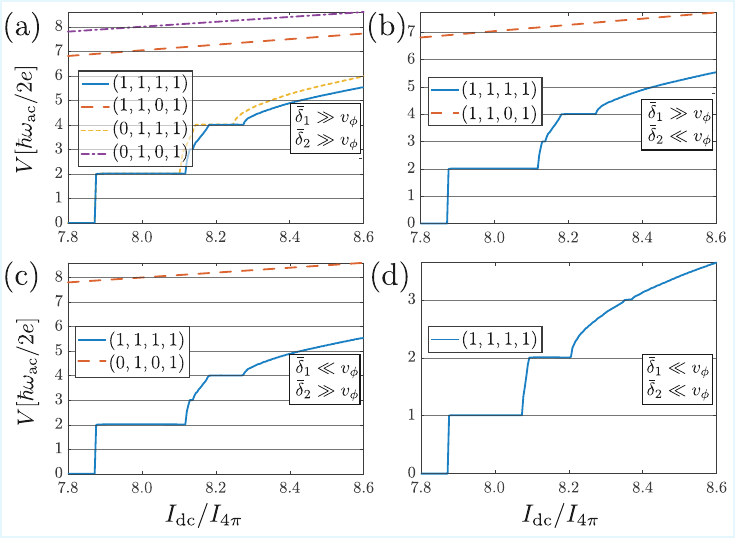}
  \caption{The voltage $V$ as a function of an applied $I_{\text{dc}}$ for different parity configurations $(n_{2\pi}^{\text{bot}},n_{2\pi}^{\text{top}},n_{4\pi}^{\text{bot}},n_{4\pi}^{\text{top}})$ and different gaps $\bar\delta_{1,2}$.
}
  \label{fig:PureVoltes}
\end{figure}

Starting with the scenario where both gaps are sufficiently large compared to the phase velocity ($\bar\delta_1=\bar\delta_2\gg v_\phi$) such that the ABS keep their individual parity fixed throughout the entire phase evolution, we see four distinct voltage curves for the different initial parity configurations, shown in Fig.~\ref{fig:PureVoltes}(a). Although there are sixteen parity configurations in total, some of them are physically equivalent. For example, the configuration $(1,1,1,0)$ and $(1,1,0,1)$ yield an identical current, since in both cases the $4\pi$-periodic current vanishes and the $2\pi$-periodic current remains. 
The four different voltage curves correspond to the four possible combinations of $2\pi$- and $4\pi$-periodic supercurrent components: one case in which both contributions are present, e.g.\ $(1,1,1,1)$; one in which only the $4\pi$-periodic current survives, e.g.\ $(0,1,1,1)$; one in which only the $2\pi$-periodic current remains, e.g.\ $(1,1,0,1)$; and one in which both contributions cancel, e.g.\ $(0,1,0,1)$. From this we can directly explain the qualitative differences between the curves. When both currents are present, we recover the usual even-integer Shapiro steps discussed in the main text. By removing the $2\pi$-component, e.g. $(0,1,1,1)$ we still see the even-integer Shapiro steps, but the plateaus are slightly shifted~\cite{Dominguez2017a}. This is because the $2\pi$-periodic current influences the time the phase spends in the even- and odd-sectors, see the discussion of Sec.~\ref{App.RSJ}. Note also that the critical current does not change in this case, since it is entirely set by the maximum value of $I_{4\pi}(\phi)$ in Eq.~(\ref{eq:4piCurrent}). In contrast, removing the $4\pi$-periodic current, as in $(1,1,0,1)$, reduces the critical current from $I_{c}/I_{4\pi}=8.0$ to $I_{c}/I_{4\pi}=4.3$, such that the onset of finite voltage occurs at smaller values of $I_{\text{dc}}$. In this case, the Shapiro steps appear at all integers, since only the $2\pi$-periodic component is present.
Finally, when both supercurrent components cancel, as in $(0,1,0,1)$, the critical current vanishes $I_{c}/I_{4\pi}=0$ and the junction simply follows Ohm's law.

In the limit where the gap between the $4\pi$-periodic ABS and the $2\pi$-periodic is large enough to suppress the exchange of parities ($\bar\delta_1\gg v_\phi$), but particles can leak from the $2\pi$-periodic ABS into the quasiparticle continuum ($\bar\delta_2\ll v_\phi$), the number of distinct voltage curves reduces from four to two, as shown in Fig.~\ref{fig:PureVoltes}(b). The reason for this reduction is that the initial values of $n_{2\pi}^{\mathrm{bot}}$ and $n_{2\pi}^{\mathrm{top}}$ do not influence the final result. After the first crossing with the quasiparticle continuum at $\phi=2\pi$, the corresponding $2\pi$-periodic ABS either remains occupied if it was already filled or becomes occupied if it was initially empty, see Figs.~\ref{fig:ABS_RSJ}(c) and (e) at $\phi=2\pi$, respectively. Consequently, after the first crossing the system always ends up in the state $(1,1,n_{4\pi}^{\text{bot}},n_{4\pi}^{\text{top}})$. Thus, we only obtain two different voltage curves: either Shapiro steps at even integers when the $4\pi$-component is present or Shapiro steps at all integers and at a reduced critical current, when the $4\pi$-component is absent.

By suppressing leakage into the quasiparticle continuum ($\bar\delta_2\gg v_\phi$) while allowing parity exchange between the $2\pi$- and $4\pi$-periodic ABS ($\bar\delta_1\ll v_\phi$), we again find only two distinct voltage responses: either even-integer Shapiro steps or Ohm’s law, see Fig.~\ref{fig:PureVoltes}(c). The configuration $(1,1,1,1)$ shows the same behavior as discussed in the main text, where the constant exchange of parities between the $4\pi$- and $2\pi$-periodic ABS leads to an effective $4\pi$-periodic response, see also Fig.~\ref{fig:ABS_RSJ}(d). Likewise, the configuration $(0,1,0,1)$ yields no supercurrent, since the exchange of parities between the two ABS always leads to a cancellation of their currents.
An interesting and rather unexpected feature appears in this parameter regime for the configurations $(1,1,0,1)$ and $(0,1,1,1)$. At first sight one might expect these two to be equivalent, because the ABS can exchange parity freely and therefore the system can transition between them at an avoided crossing. However, the configuration $(1,1,0,1)$ exhibits even-integer Shapiro steps, whereas $(0,1,1,1)$ shows Ohm’s law.
This difference originates from the form of the ABS, which alternate between “curved’’ and “flat’’ regions in the energy–phase relation, shown in Fig.~\ref{fig:ABS_RSJ}(a). When one ABS exhibits curvature in the phase $\phi$ and carries supercurrent, the other one is flat and contributes essentially nothing. In the configuration $(1,1,0,1)$, the cancellation of the $4\pi$-currents occurs precisely in a phase region where the corresponding $4\pi$-periodic ABS is flat and would not contribute to the current anyway. At the first avoided crossing, the system transitions to $(0,1,1,1)$, where now the $2\pi$-current cancels, again in a region where the $2\pi$-ABS is flat and does not carry current. Thus, throughout the phase evolution the system always cancels the ABS that is \emph{not} contributing and preserves the one that \emph{is}. As a result, the dynamics of $(1,1,0,1)$ is effectively identical to $(1,1,1,1)$ and both yield even-integer Shapiro steps.
In contrast, the configuration $(0,1,1,1)$ cancels precisely the ABS that \emph{does} carry current, while keeping the flat, non–current-carrying ABS intact. Consequently, the supercurrent is suppressed at every value of the phase $\phi$, and the system behaves equivalently to $(0,1,0,1)$, yielding Ohm’s law.

Finally, in the limit where both gaps are negligible compared to $v_\phi$ ($\bar\delta_1=\bar\delta_2\ll v_\phi$), all parity configurations yield the exact same voltage, shown in Fig.~\ref{fig:PureVoltes}(d). Independently of the initial configuration, the spectrum repeats itself at every $\phi=2\pi n$, $n\in\mathbb{Z}$, due to the coupling to the quasipaticle continuum. As a consequence, all parity configurations yield Shapiro steps at all integers.

\subsection{ ~~Dwell-time weighted average}
\label{App: Dwell-Time weighted average}
\begin{figure}[tb]
  \centering
  \includegraphics[width=1\linewidth]{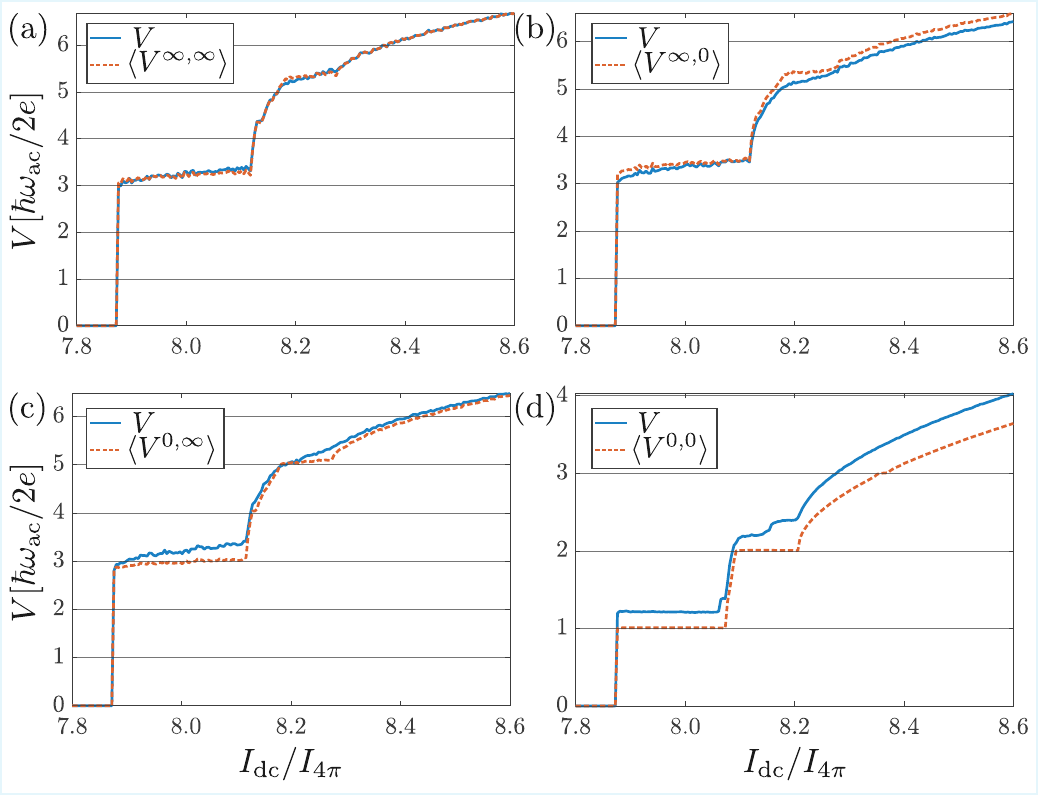}
  \caption{The dwell--time weighted average $\langle V^{\bar\delta_1^\prime,\bar\delta_2^\prime}\rangle$ for different values of $\bar\delta_{1/2}^\prime$, compared the voltage of the full RSJ calculation, also shown in Fig.~\ref{fig:VoltageCurves}. Panel (a)-(f) correspond to Fig.~\ref{fig:VoltageCurves}(i), (e), (f), (j), (l) and (m), respectively. In this notation the superindex $\infty$ means $\bar\delta\gg v_\phi$ and $0$ means $\bar\delta\ll v_\phi$. 
}\label{fig:DwellTimeWeightedAverage}
\end{figure}
In this subsection, we show that we can approximate the voltage of the full RSJ dynamics by averaging the voltages of the individual parity configurations of App.~\ref{Parity configuration}, weighted by the time the system spends in each configuration. We refer to this as the dwell-time weighted average. This method provides a simple qualitative explanation for the origin of the non-quantized steps, which we discussed in the main text when the gaps are in the regime of $\bar\delta_{1/2}\approx v_\phi$, as illustrated in Fig.~\ref{fig:VoltageCurves}(f), (g), (j), (k), (n) and (o). 

\begin{figure}[tb]
  \centering
  \includegraphics[width=1\linewidth]{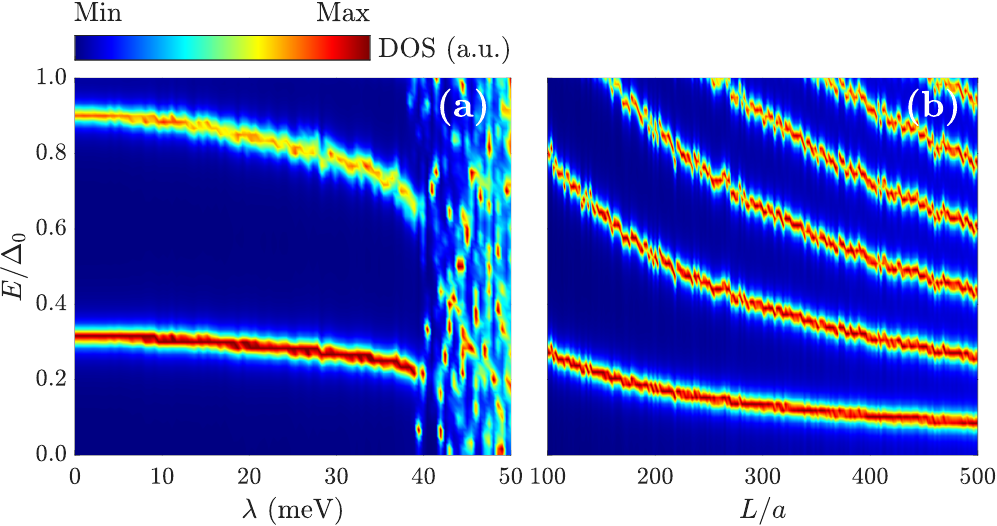}
  \caption{Panel~(a): The energy of the phase-independent ABS as a function of the disorder strength $\lambda$. We set $L_S=\infty$ and the length and width of the normal part to $L=W=0.5$~$\mu$m, respectively. In this plot, we averaged over $20$ disorder configurations.  Panel~(b): The energy of the phase-independent ABS as a function of the length of the normal part $L$, with $a$ as the lattice constant introduced in the main part of the text. We set $\lambda=35$~meV. }
  \label{fig:Disorder}
\end{figure}\noindent

The idea is the following: in the full RSJ calculation we let the system start in a parity configuration $n=(n_{2\pi}^{\mathrm{bot}}, n_{2\pi}^{\mathrm{top}}, n_{4\pi}^{\mathrm{bot}}, n_{4\pi}^{\mathrm{top}})$ and depending on $\bar\delta_1$ and $\bar\delta_2$ Landau--Zener transitions occur which change this initial parity configuration. The system then spends a time $\tau_n$ in this parity configuration before undergoing another LZ transition. To approximate the generated voltage of this situation, we take the following average
\begin{align}
\label{DwellTimeWeightedAverage}
    \langle V^{\bar\delta_1^\prime,\bar\delta_2^\prime} \rangle=\sum_n\frac{\tau_n V_n(\bar\delta_1^\prime,\bar\delta_2^\prime)}{\sum_n\tau_n},
\end{align}
where $V_n(\bar\delta_1^\prime,\bar\delta_2^\prime)$ is the voltage for a given parity configuration $n$ and for given gaps $\bar\delta_1^\prime$ and $\bar\delta_2^\prime$, i.e. the curves shown in Fig.~\ref{fig:PureVoltes}. Note that the gaps $\bar\delta_{1,2}^\prime$ in Eq.~(\ref{DwellTimeWeightedAverage}) are always in the limit of $\bar\delta_{1/2}^\prime \mathrel{\substack{\ll\\\gg}} v_\phi$, because they refer to the voltage curves discussed in App.~\ref{Parity configuration}. In contrast, the gaps $\bar\delta_{1/2}$ of the full RSJ calculation of the main text can take any value. As a result, Eq.~(\ref{DwellTimeWeightedAverage}) can only approximate the full RSJ calculation. However, this method captures the qualitative behaviour of the non-quantized steps, as illustrated in Fig.~\ref{fig:DwellTimeWeightedAverage} and therefore provides an intuitive picture of how the LZ transitions between different parity states lead to voltage plateaus at non-integer values.

  \begin{figure}[t]
  \centering
  \includegraphics[width=6.3in]{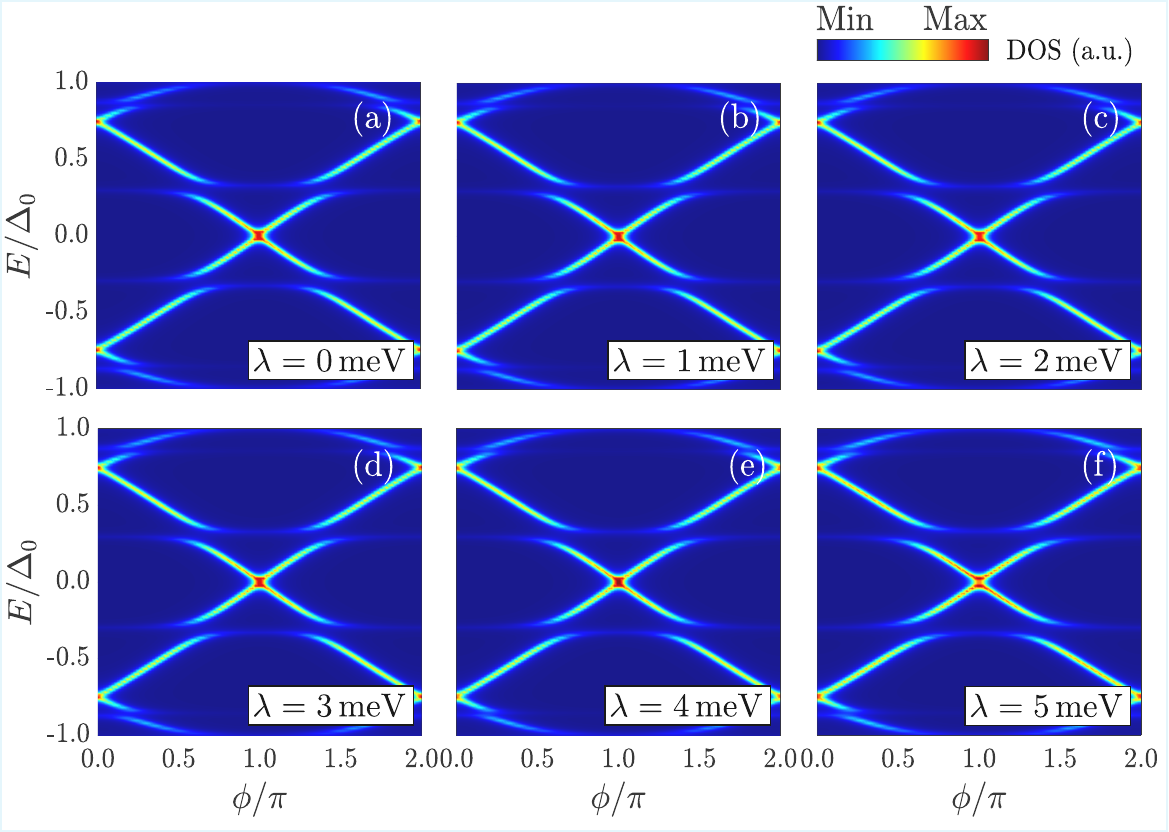}
  \caption{The ABS calculated for different on-site disorder strengths $\lambda$ in the superconducting regions. In contrast to the main text, we use $C_n=-6\,$meV and $L_{N'}=0.35\,$$\mu$m.}
  \label{fig:DisorderABS}
\end{figure}
\section{ ~~Disorder effects}
\label{App.disorder}
To investigate the effects of disorder on the transport properties of the Josephson junction, we add  $\mathcal{H}_{Dis}=\lambda s \sigma_0 \otimes \tau _z$ to the Hamiltonian $\mathcal{H}$ of Eq.~(\ref{eq:BdGBHZ}), where $\lambda$ is the disorder strength, $\tau$ represents the particle-hole degree of freedom and $s$ takes on random values between $-1$ and $1$ for each lattice site. Here, we restrict the analysis to a single external part attached to a semi-infinite superconductor~\cite{Lopez1984a}. In Fig.~\ref{fig:Disorder}(a) we observe that the phase-independent ABS are stable in the presence of disorder up to a critical strength of $\lambda_c \approx 4 |M|$. Furthermore, we note that as $\lambda$ increases, the energy of the ABS diminishes, since disorder forces the particles to make a detour. Disorder strengths larger than the bulk gap can couple the top and bottom edges at every spatial point of the junction, lifting the time-reversal protection of the states.

In Fig.~\ref{fig:Disorder} (b) we show that for a high disorder strength $\lambda=35\, \text{meV}$ the ABS remain stable upon increasing the length of the normal part. Adding more sites to the QSHI allows for more scattering events of the helical edge with disordered sites, however, as long as the disorder strength stays below the critical value of $\lambda_c$ the ABS obey Eq.~(\ref{eq:analyticalresult}) up to minor distortions, resulting from a slightly altered perimeter. As previously mentioned, we also observe that the disorder has a progressively greater impact on the higher ABS.  
Returning to the full N'SNSN' junction, we now add the on-site disorder potential only in the superconducting regions. For the numerical analysis we set up the Josephson junction with $L_s = 0.5\,\mu$m and $\xi_s \sim 200\,$nm, and include disorder strengths $\lambda = 0$–$5\,$meV. We find that disorder affects the avoided level crossing only slightly, and in a nonmonotonic way, leading to either an increase or a decrease of the gap size as the disorder strength increases, see Fig.~\ref{fig:DisorderABS}. Therefore, we expect to observe the isolation of a $4\pi$-periodic ABS for a wide range of disorder strengths.
\section{ ~~Direct edge coupling}
\label{App:Direct edge coupling}
\begin{figure}[t]
  \centering
  \includegraphics[width=6.3in]{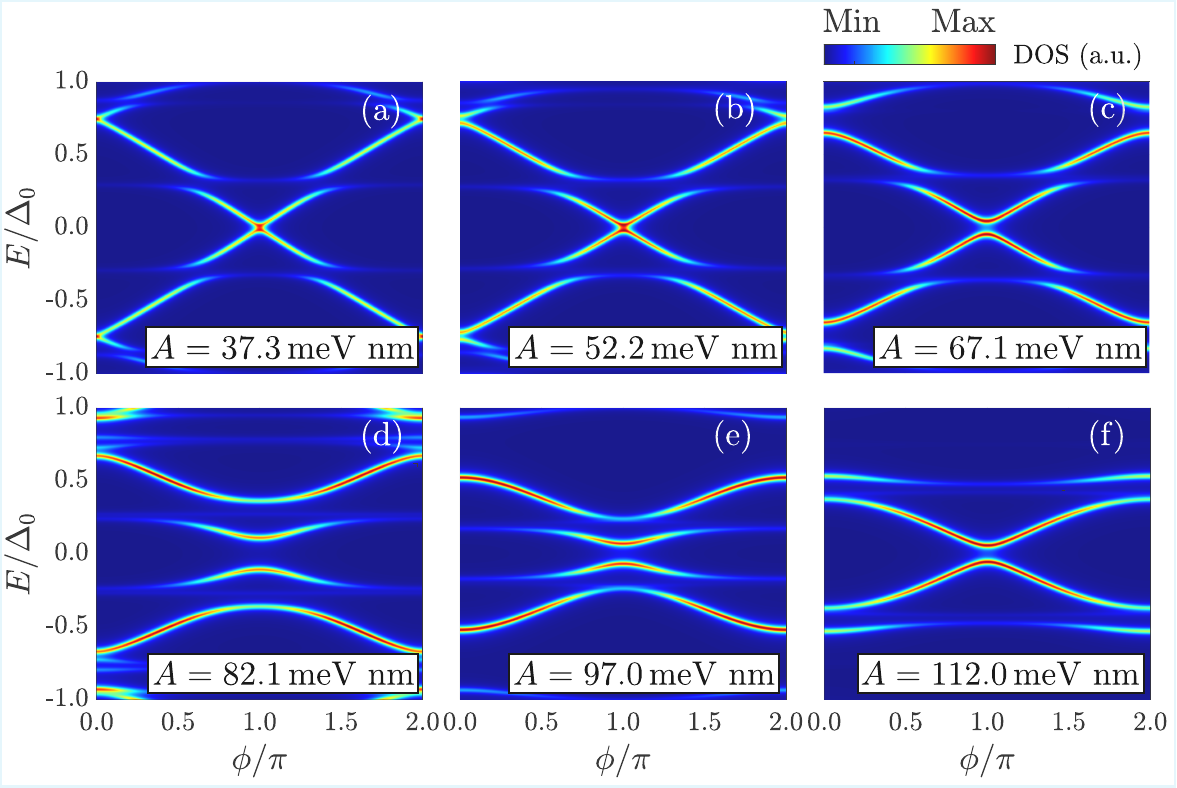}
  \caption{Simulation of band bending effects showing the ABS for different values of the $A$ parameter, which controls the Fermi velocity in the superconducting regions. In contrast to the main text, we use $C_n=-6\,$meV and $L_{N'}=0.35\,\mu$m.}
  \label{fig:DifferentAs}
\end{figure}
The presence of doping and band-bending is a typical consequence of coupling a QSHI to an external superconductor. Both of these effects can enhance a direct interedge backscattering process along the NS interface. As a result of this direct coupling the ABS develop anticrossings at  $\phi=n\pi$, with $n\in \mathbb{Z}$, destroying the fractional Josephson effect. Note that this is not specific to our model but to all Josephson junctions based on quantum spin Hall systems. In practice, there is a broad parameter range in which the top and bottom QSHI edges behave as two independent junctions, with the same physics as in the Fu and Kane model~\cite{Fu2009a}.

More specifically, two distinct effects enhance the direct interedge coupling mechanism: (i) When the chemical potential in the superconducting regions lies within the bulk bands, the top and bottom edge states of the QSHI can couple through the bulk~\cite{Recher2013}. This bulk-mediated edge coupling can destroy the fractional Josephson effect by opening a gap in the ABS at zero energy. The strength of the coupling is determined by the doping level in the superconducting regions and the spatial separation between the edges.
(ii) Band bending effects can change the Fermi velocity of the proximitized regions, reducing the effective coupling between the QSH edges and the superconductor.

When only one of these effects is present, the top and bottom edge states are typically decoupled because they are far apart from each other. As a numerical example, note that we have used highly doped superconducting regions in the main text, where we do not observe a sizable gap opening at zero energy. On the other hand, band bending on proximitized helical edges does not affect the coupling to the QSH edge state, since there is essentially no other state the quasiparticles can scatter into. Thus, the only possible processes are local Andreev reflection or electron-electron tunneling through the superconductor on a single edge.

\begin{figure}[t]
  \centering
  \includegraphics[width=6.3in]{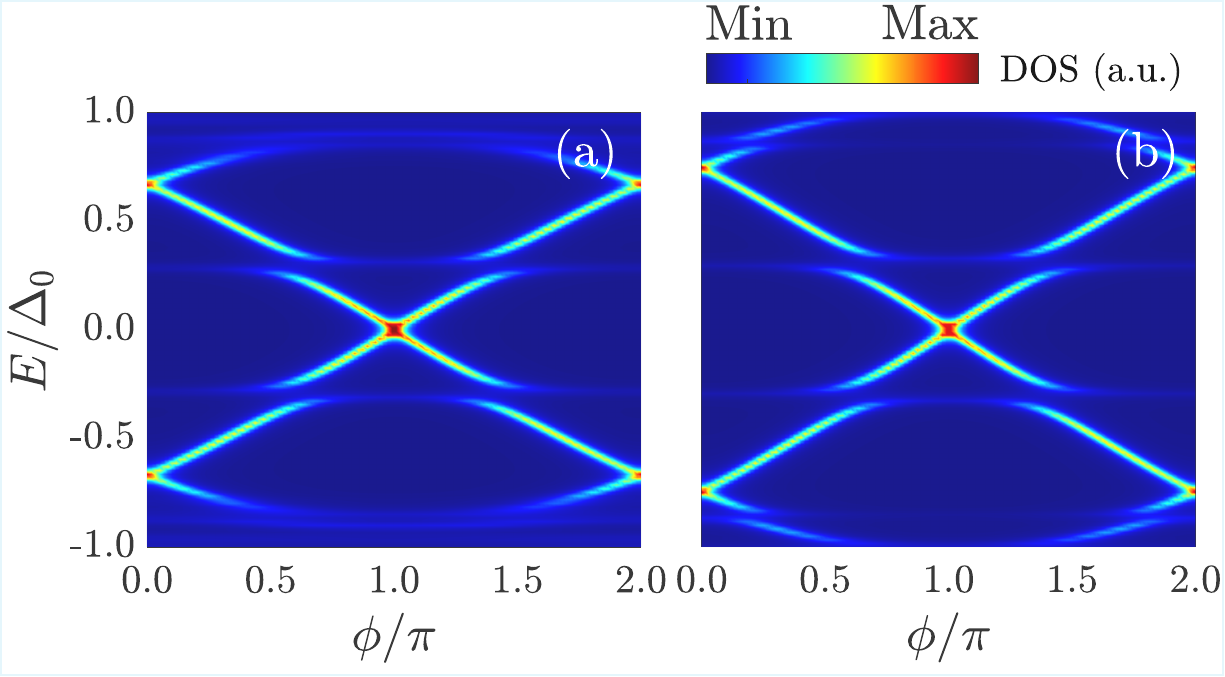}
  \caption{(a) The ABS calculated with an $\omega$-dependent self-energy, with a parent gap of $\Delta=3\,$meV and an induced gap of $\Delta_{ind}=\Delta_0=0.6\,$meV. (b) The ABS calculated in the same way as in the main text, but with  $C_n=-6\,$meV. In contrast to the main text, we use $C_n=-6\,$meV and $L_{N'}=0.35\,$$\mu$m in (a) and (b).}
  \label{fig:SCSelfEnergy}
\end{figure}

The presence of band bending effects and doping enhances the interedge coupling through the NS interfaces mediated by the transversal modes generated along them. We can see in Fig.~\ref{fig:DifferentAs} a set of ABS corresponding to a N'SNSN' setup with highly doped proximitized regions and different ratios $v_{F,s}/v_{F,n}$, set by a BHZ parameter $A$ on the proximitized regions. We can observe that, already, for a ratio of $v_{F,s}/v_{F,n}=2$, a gap at zero energy opens, which spoils the fractional Josephson effect. Besides, the lack of transparency reduces the coupling between the N and N' regions, reducing the coupling between phase-dependent and independent ABS.

In summary, our decoupling scheme relies on similar conditions as the fractional Josephson effect, that is, transparent NS interfaces to avoid a direct interedge coupling. These effects can be reduced by an appropriate choice of superconductor and geometry of the junction.

  \begin{figure}[t]
  \centering
  \includegraphics[width=6.3in]{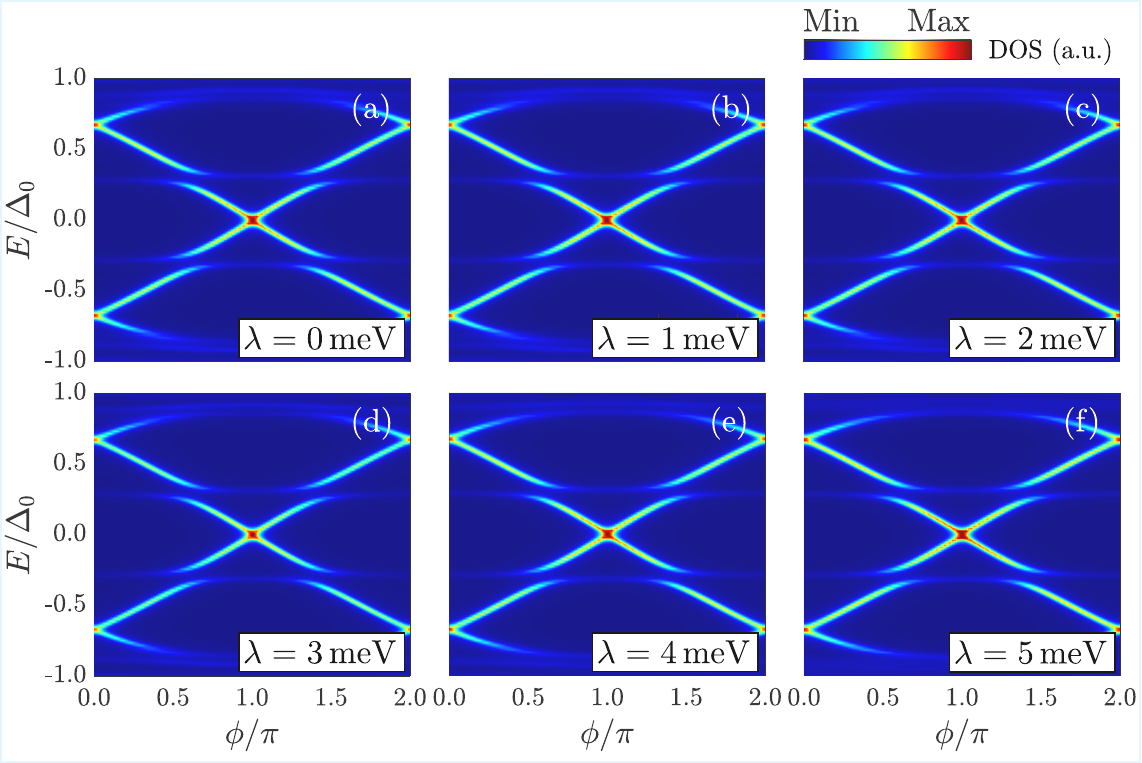}
  \caption{The ABS calculated with an $\omega$-dependent self-energy and for different on-site disorder strengths $\lambda$ in the superconducting regions. In contrast to the main text, we use $C_n=-6\,$meV and $L_{N'}=0.35\,$$\mu$m.}
  \label{fig:Disorder+Selfenergy}
\end{figure}

\section{ ~~Energy-independent superconducting pairing}
\label{App: Energy-independent superconducting pairing}

In Sec.~\ref{Tight-binding model} of the main text, we use a standard approximation in which the proximitized region inherits an energy-independent pairing potential from the parent superconductor. This approximation is justified when the induced pairing $\Delta_{\mathrm{ind}}$ is much smaller than the gap $\Delta$ of the parent superconductor, i.e.\ $\Delta_{\mathrm{ind}} \ll \Delta$~\cite{Cole2016,Stanescu2017}.
To see more explicitly how this approximation works, we couple a single BHZ site to an s-wave superconductor and integrate out the parent superconductor. The resulting $\omega$-dependent self-energy then takes the following form
\begin{align}
    \Sigma_{SC}=-\frac{\Delta_{ind}}{\sqrt{\Delta^2-\omega^2}}\begin{pmatrix}
        \omega & 0 & \Delta & 0 \\
        0 &  \omega & 0 &  \Delta \\
        \Delta  & 0 &  \omega & 0 \\
        0 & \Delta& 0 &  \omega \\
    \end{pmatrix},
\end{align}
with $\Delta_{ind}=\pi \rho_s t^2$, $\rho_s$ as the local density of states of the superconductor and $t$ the tunneling amplitude to the superconductor.
The self-energy contains two distinct local terms within the proximitized regions, one anomalous $\propto \Delta/\sqrt{\Delta^2-\omega^2} $ and a broadening or a life-time $\propto \omega /\sqrt{\Delta^2-\omega^2}$. In order to investigate the effects of the $\omega$-dependent self-energy numerically, we set a parent gap of $\Delta=3\,$meV and  an induced gap of $\Delta_{ind}=0.6\,$meV, shown in Fig.~\ref{fig:SCSelfEnergy} (b). Comparing the scenario of the $\omega$-dependent self-energy to the $\omega$-independent self-energy of the main text, shown in Fig.~\ref{fig:SCSelfEnergy} (a), we find no qualitative differences for lowest-lying avoided level crossing.

In the diffusive limit, coupling to the superconductor induces disorder in the proximitized regions, either through a disordered potential or through disordered pairing. To investigate these effects, we include on-site disorder in the superconducting region, following App.~\ref{App.disorder}, but we find no qualitative change in the ABS, as shown in Fig.~\ref{fig:Disorder+Selfenergy}.

\end{appendix}

\bibliography{bibliography}

\begin{thebibliography}{10}
\providecommand{\url}[1]{\texttt{#1}}
\providecommand{\urlprefix}{URL }
\expandafter\ifx\csname urlstyle\endcsname\relax
  \providecommand{\doi}[1]{doi:\discretionary{}{}{}#1}\else
  \providecommand{\doi}{doi:\discretionary{}{}{}\begingroup
  \urlstyle{rm}\Url}\fi
\providecommand{\eprint}[2][]{\url{#2}}

\bibitem{Fu2009a}
L.~Fu and C.~L. Kane,
\newblock \emph{Josephson current and noise at a
  superconductor/quantum-spin-hall-insulator/superconductor junction},
\newblock Phys. Rev. B \textbf{79}, 161408 (2009),
\newblock \doi{10.1103/PhysRevB.79.161408}.

\bibitem{Beenakker2013a}
C.~W.~J. Beenakker, D.~I. Pikulin, T.~Hyart, H.~Schomerus and J.~P. Dahlhaus,
\newblock \emph{Fermion-parity anomaly of the critical supercurrent in the
  quantum spin-hall effect},
\newblock Phys. Rev. Lett. \textbf{110}, 017003 (2013),
\newblock \doi{10.1103/PhysRevLett.110.017003}.

\bibitem{Hart2014}
S.~Hart, H.~Ren, T.~Wagner, P.~Leubner, M.~Mühlbauer, C.~Brüne, H.~Buhmann,
  L.~W. Molenkamp and A.~Yacoby,
\newblock \emph{Induced superconductivity in the quantum spin hall edge},
\newblock Nat. Phys. \textbf{10}, 638 (2014),
\newblock \doi{10.1038/nphys3036}.

\bibitem{Pribiag2015a}
V.~S. Pribiag, A.~J.~A. Beukman, F.~Qu, M.~C. Cassidy, C.~Charpentier,
  W.~Wegscheider and L.~P. Kouwenhoven,
\newblock \emph{Edge-mode superconductivity in a two-dimensional topological
  insulator},
\newblock Nat. Nanotech. \textbf{10}, 593 (2015),
\newblock \doi{10.1038/nnano.2015.86}.

\bibitem{Bocquillon2016a}
E.~Bocquillon, R.~S. Deacon, J.~Wiedenmann, L.~P., T.~M. Klapwijk, C.~Br\"une,
  K.~Ishibashi, H.~Buhmann and L.~W. Molenkamp,
\newblock \emph{Gapless andreev bound states in the quantum spin hall insulator
  hgte},
\newblock Nat. Nano. \textbf{12}, 137 (2016),
\newblock \doi{10.1038/nnano.2016.159}.

\bibitem{Deacon2017a}
R.~S. Deacon, J.~Wiedenmann, E.~Bocquillon, F.~Dom\'{\i}nguez, T.~M. Klapwijk,
  P.~Leubner, C.~Br\"une, E.~M. Hankiewicz, S.~Tarucha, K.~Ishibashi,
  H.~Buhmann and L.~W. Molenkamp,
\newblock \emph{Josephson radiation from gapless andreev bound states in
  hgte-based topological junctions},
\newblock Phys. Rev. X \textbf{7}, 021011 (2017),
\newblock \doi{10.1103/PhysRevX.7.021011}.

\bibitem{Bendias2018}
K.~Bendias, S.~Shamim, O.~Herrmann, A.~Budewitz, P.~Shekhar, P.~Leubner,
  J.~Kleinlein, E.~Bocquillon, H.~Buhmann and L.~W. Molenkamp,
\newblock \emph{High mobility hgte microstructures for quantum spin hall
  studies},
\newblock Nano Lett. \textbf{18}, 4831 (2018),
\newblock \doi{0.1002/adma.202301683}.

\bibitem{Randle2023}
M.~D. Randle, M.~Hosoda, R.~S. Deacon, M.~Ohtomo, P.~Zellekens, K.~Watanabe,
  T.~Taniguchi, S.~Okazaki, T.~Sasagawa, K.~Kawaguchi, S.~Sato and
  K.~Ishibashi,
\newblock \emph{Gate-defined josephson weak-links in monolayer wte2},
\newblock Advanced Materials \textbf{35}(35), 2301683 (2023),
\newblock \doi{https://doi.org/10.1002/adma.202301683},
\newblock
  \eprint{https://onlinelibrary.wiley.com/doi/pdf/10.1002/adma.202301683}.

\bibitem{Sticlet2018a}
D.~Sticlet, J.~D. Sau and A.~Akhmerov,
\newblock \emph{Dissipation-enabled fractional josephson effect},
\newblock Phys. Rev. B \textbf{98}, 125124 (2018),
\newblock \doi{10.1103/PhysRevB.98.125124}.

\bibitem{Lahiri2023a}
A.~Lahiri, S.-J. Choi and B.~Trauzettel,
\newblock \emph{Nonequilibrium fractional josephson effect},
\newblock Phys. Rev. Lett. \textbf{131}, 126301 (2023),
\newblock \doi{10.1103/PhysRevLett.131.126301}.

\bibitem{Yeyati2003a}
A.~L. Yeyati, A.~Mart\'{\i}n-Rodero and E.~Vecino,
\newblock \emph{Nonequilibrium dynamics of andreev states in the kondo regime},
\newblock Phys. Rev. Lett. \textbf{91}, 266802 (2003),
\newblock \doi{10.1103/PhysRevLett.91.266802}.

\bibitem{San-Jose2012a}
P.~San-Jose, E.~Prada and R.~Aguado,
\newblock \emph{ac josephson effect in finite-length nanowire junctions with
  majorana modes},
\newblock Phys. Rev. Lett. \textbf{108}, 257001 (2012),
\newblock \doi{10.1103/PhysRevLett.108.257001}.

\bibitem{Dominguez2012a}
F.~Dom\'{\i}nguez, F.~Hassler and G.~Platero,
\newblock \emph{Dynamical detection of majorana fermions in current-biased
  nanowires},
\newblock Phys. Rev. B \textbf{86}, 140503 (2012),
\newblock \doi{10.1103/PhysRevB.86.140503}.

\bibitem{Pikulin2012a}
D.~I. Pikulin and Y.~V. Nazarov,
\newblock \emph{Phenomenology and dynamics of a majorana josephson junction},
\newblock Phys. Rev. B \textbf{86}, 140504 (2012),
\newblock \doi{10.1103/PhysRevB.86.140504}.

\bibitem{Houzet2013a}
M.~Houzet, J.~S. Meyer, D.~M. Badiane and L.~I. Glazman,
\newblock \emph{Dynamics of majorana states in a topological josephson
  junction},
\newblock Phys. Rev. Lett. \textbf{111}, 046401 (2013),
\newblock \doi{10.1103/PhysRevLett.111.046401}.

\bibitem{Virtanen2013a}
P.~Virtanen and P.~Recher,
\newblock \emph{Microwave spectroscopy of josephson junctions in topological
  superconductors},
\newblock Phys. Rev. B \textbf{88}, 144507 (2013),
\newblock \doi{10.1103/PhysRevB.88.144507}.

\bibitem{Matthews2014a}
P.~Matthews, P.~Ribeiro and A.~M. Garc\'{\i}a-Garc\'{\i}a,
\newblock \emph{Dissipation in a simple model of a topological josephson
  junction},
\newblock Phys. Rev. Lett. \textbf{112}, 247001 (2014),
\newblock \doi{10.1103/PhysRevLett.112.247001}.

\bibitem{Sau2017a}
J.~D. Sau and F.~Setiawan,
\newblock \emph{Detecting topological superconductivity using low-frequency
  doubled shapiro steps},
\newblock Phys. Rev. B \textbf{95}, 060501 (2017),
\newblock \doi{10.1103/PhysRevB.95.060501}.

\bibitem{liu2024}
W.~Liu, S.~U. Piatrusha, X.~Liang, S.~Upadhyay, L.~Fürst, C.~Gould,
  J.~Kleinlein, H.~Buhmann, M.~P. Stehno and L.~W. Molenkamp,
\newblock \emph{Period-doubling in the phase dynamics of a shunted hgte quantum
  well josephson junction},
\newblock Nat Commun \textbf{16}, 3068 (2025),
\newblock \doi{10.1103/PhysRevB.79.161408}.

\bibitem{Bernevig2006a}
B.~A. Bernevig, T.~L. Hughes and S.-C. Zhang,
\newblock \emph{Quantum spin hall effect and topological phase transition in
  hgte quantum wells},
\newblock Science \textbf{314}(5806), 1757 (2006),
\newblock \doi{10.1126/science.1133734},
\newblock \eprint{https://www.science.org/doi/pdf/10.1126/science.1133734}.

\bibitem{Liu2008a}
C.~Liu, T.~L. Hughes, X.-L. Qi, K.~Wang and S.-C. Zhang,
\newblock \emph{Quantum spin hall effect in inverted type-ii semiconductors},
\newblock Phys. Rev. Lett. \textbf{100}, 236601 (2008),
\newblock \doi{10.1103/PhysRevLett.100.236601}.

\bibitem{Murani2017a}
A.~Murani, A.~Chepelianskii, S.~Gu\'eron and H.~Bouchiat,
\newblock \emph{Andreev spectrum with high spin-orbit interactions: Revealing
  spin splitting and topologically protected crossings},
\newblock Phys. Rev. B \textbf{96}, 165415 (2017),
\newblock \doi{10.1103/PhysRevB.96.165415}.

\bibitem{Zhou2008a}
B.~Zhou, H.-Z. Lu, R.-L. Chu, S.-Q. Shen and Q.~Niu,
\newblock \emph{Finite size effects on helical edge states in a quantum
  spin-hall system},
\newblock Phys. Rev. Lett. \textbf{101}, 246807 (2008),
\newblock \doi{10.1103/PhysRevLett.101.246807}.

\bibitem{Reinthaler2013a}
R.~W. Reinthaler, P.~Recher and E.~M. Hankiewicz,
\newblock \emph{Proposal for an all-electrical detection of crossed andreev
  reflection in topological insulators},
\newblock Phys. Rev. Lett. \textbf{110}, 226802 (2013),
\newblock \doi{10.1103/PhysRevLett.110.226802}.

\bibitem{MacKinnon1985}
A.~MacKinnon,
\newblock \emph{The calculation of transport properties and density of states
  of disordered solids},
\newblock Zeitschrift für Physik B Condensed Matter \textbf{59}, 385 (1985),
\newblock \doi{10.1007/BF01328846}.

\bibitem{Finocchiaro2018a}
F.~Finocchiaro, F.~Guinea and P.~San-Jose,
\newblock \emph{Topological $\ensuremath{\pi}$ junctions from crossed andreev
  reflection in the quantum hall regime},
\newblock Phys. Rev. Lett. \textbf{120}, 116801 (2018),
\newblock \doi{10.1103/PhysRevLett.120.116801}.

\bibitem{Knapp2020a}
C.~Knapp, A.~Chew and J.~Alicea,
\newblock \emph{Fragility of the fractional josephson effect in
  time-reversal-invariant topological superconductors},
\newblock Phys. Rev. Lett. \textbf{125}, 207002 (2020),
\newblock \doi{10.1103/PhysRevLett.125.207002}.

\bibitem{Lee2014a}
S.-P. Lee, K.~Michaeli, J.~Alicea and A.~Yacoby,
\newblock \emph{Revealing topological superconductivity in extended quantum
  spin hall josephson junctions},
\newblock Phys. Rev. Lett. \textbf{113}, 197001 (2014),
\newblock \doi{10.1103/PhysRevLett.113.197001}.

\bibitem{Dominguez2024a}
F.~Dominguez, E.~G. Novik and P.~Recher,
\newblock \emph{Fraunhofer pattern in the presence of majorana zero modes},
\newblock Phys. Rev. Res. \textbf{6}, 023304 (2024),
\newblock \doi{10.1103/PhysRevResearch.6.023304}.

\bibitem{Recher2013}
R.~W. Reinthaler, P.~Recher and E.~M. Hankiewicz,
\newblock \emph{Proposal for an all-electrical detection of crossed andreev
  reflection in topological insulators},
\newblock Phys. Rev. Lett. \textbf{110}, 226802 (2013),
\newblock \doi{10.1103/PhysRevLett.110.226802}.

\bibitem{Dominguez2017a}
F.~Dom\'{\i}nguez, O.~Kashuba, E.~Bocquillon, J.~Wiedenmann, R.~S. Deacon,
  T.~M. Klapwijk, G.~Platero, L.~W. Molenkamp, B.~Trauzettel and E.~M.
  Hankiewicz,
\newblock \emph{Josephson junction dynamics in the presence of
  $2\ensuremath{\pi}$- and $4\ensuremath{\pi}$-periodic supercurrents},
\newblock Phys. Rev. B \textbf{95}, 195430 (2017),
\newblock \doi{10.1103/PhysRevB.95.195430}.

\bibitem{Dynes1971a}
R.~C. Dynes and T.~A. Fulton,
\newblock \emph{Supercurrent density distribution in josephson junctions},
\newblock Phys. Rev. B \textbf{3}, 3015 (1971),
\newblock \doi{10.1103/PhysRevB.3.3015}.

\bibitem{Baxevanis2015a}
B.~Baxevanis, V.~P. Ostroukh and C.~W.~J. Beenakker,
\newblock \emph{Even-odd flux quanta effect in the fraunhofer oscillations of
  an edge-channel josephson junction},
\newblock Phys. Rev. B \textbf{91}, 041409 (2015),
\newblock \doi{10.1103/PhysRevB.91.041409}.

\bibitem{Crepin2014a}
F.~Cr\'epin and B.~Trauzettel,
\newblock \emph{Parity measurement in topological josephson junctions},
\newblock Phys. Rev. Lett. \textbf{112}, 077002 (2014),
\newblock \doi{10.1103/PhysRevLett.112.077002}.

\bibitem{Rainis2012}
D.~Rainis and D.~Loss,
\newblock \emph{Majorana qubit decoherence by quasiparticle poisoning},
\newblock Phys. Rev. B \textbf{85}, 174533 (2012),
\newblock \doi{https://doi.org/10.1103/PhysRevB.85.174533}.

\bibitem{Datta1995a}
S.~Datta,
\newblock \emph{Electronic Transport in Mesoscopic Systems},
\newblock Cambridge University Press, Cambridge (1995).

\bibitem{Tkachov2015a}
G.~Tkachov, P.~Burset, B.~Trauzettel and E.~M. Hankiewicz,
\newblock \emph{Quantum interference of edge supercurrents in a two-dimensional
  topological insulator},
\newblock Phys. Rev. B \textbf{92}, 045408 (2015),
\newblock \doi{10.1103/PhysRevB.92.045408}.

\bibitem{Haidekker2020a}
T.~Haidekker~Galambos, S.~Hoffman, P.~Recher, J.~Klinovaja and D.~Loss,
\newblock \emph{Superconducting quantum interference in edge state josephson
  junctions},
\newblock Phys. Rev. Lett. \textbf{125}, 157701 (2020),
\newblock \doi{10.1103/PhysRevLett.125.157701}.

\bibitem{Vigliotti_2022}
L.~Vigliotti, A.~Calzona, B.~Trauzettel, M.~Sassetti and N.~T. Ziani,
\newblock \emph{Anomalous flux periodicity in proximitised quantum spin hall
  constrictions},
\newblock New Journal of Physics \textbf{24}(5), 053017 (2022),
\newblock \doi{10.1088/1367-2630/ac643b}.

\bibitem{Viglioti2023a}
L.~Vigliotti, A.~Calzona, N.~Traverso~Ziani, F.~S. Bergeret, M.~Sassetti and
  B.~Trauzettel,
\newblock \emph{Effects of the spatial extension of the edge channels on the
  interference pattern of a helical josephson junction},
\newblock Nanomaterials \textbf{13}(3) (2023),
\newblock \doi{10.3390/nano13030569}.

\bibitem{Lopez1984a}
M.~P.~L. Sancho, J.~M.~L. Sancho and J.~Rubio,
\newblock \emph{Highly convergent schemes for the calculation of bulk and
  surface green functions},
\newblock Journal of Physics F: Metal Physics \textbf{15}(4), 851 (1985),
\newblock \doi{10.1088/0305-4608/15/4/009}.

\bibitem{Cole2016}
W.~S. Cole, S.~J. D.,  and S.~Das~Sarma,
\newblock \emph{Proximity effect and majorana bound states in clean
  semiconductor nanowires coupled to disordered superconductors},
\newblock Phys. Rev. B \textbf{94}, 140505(R) (2016),
\newblock \doi{https://doi.org/10.1103/PhysRevB.94.140505}.

\bibitem{Stanescu2017}
T.~D. Stanescu and S.~Das~Sarma,
\newblock \emph{Proximity-induced low-energy renormalization in hybrid
  semiconductor-superconductor majorana structures},
\newblock Phys. Rev. B \textbf{96}, 014510 (2017),
\newblock \doi{https://doi.org/10.1103/PhysRevB.96.014510}.

\end{thebibliography}

\nolinenumbers

\end{document}